\documentclass[useAMS,usenatbib]{mn2e} 
\usepackage{graphicx}
\usepackage{color}

\usepackage{txfonts}
\title[Chemical abundances of local RR Lyrae]{Chemical abundances
of solar neighborhood RR Lyrae stars\thanks{Based on data collected with
UVES@VLT under program ID 083.B-0281 and with SARG@TNG under program IDs AOT~19
TAC~11 and AOT~20 TAC~83. Also based on ESO FEROS and HARPS archival reduced
data products, under program IDs 079.D-0462 and 178.D-0361.}}

\author[E.~Pancino et al.]
{E. Pancino$^{1,2}$\thanks{email:elena.pancino@oabo.inaf.it},
N. Britavskiy$^{3,4,5}$,
D. Romano$^{1}$,
C. Cacciari$^{1}$,
A. Mucciarelli$^{5}$,
\newauthor and G. Clementini$^{1}$\\
$^{1}$INAF-Osservatorio Astronomico di Bologna, Via Ranzani 1, I-40127 Bologna,
  Italy\\
$^{2}$ASI Science Data Center, Via del Politecnico snc, I-00133 Roma, Italy\\
$^{3}$IAASARS, National Observatory of Athens, GR-15236 Penteli, Greece\\
$^{4}$Department of Astronomy and Astronomical Observatory, Odessa National
  University, T.G. Shevchenko Park, Odessa, 65014, Ukraine\\
$^{5}$Dipartimento di Fisica e Astronomia, Universit\`a di Bologna, Viale Berti
  Pichat 6/2, I-40127 Bologna, Italy}

\begin{document}

\date{Accepted ... Received ...; in original form ...}

\pagerange{\pageref{firstpage}--\pageref{lastpage}} \pubyear{2013}

\maketitle
\label{firstpage}

\begin{abstract}

We have analysed a sample of 18 RR Lyrae stars (17
fundamental-mode --- RRab --- and one first overtone --- RRc) and three
Population II Cepheids (two BL~Her stars and one W~Vir star), for which
high-resolution (R~$\ge$30\,000), high  signal-to-noise (S/N~$\ge$30) spectra
were obtained with either SARG at the Telescopio Nazionale Galileo (La Palma,
Spain) or UVES at the ESO Very Large Telescope (Paranal, Chile). Archival data
were also analyzed for a few stars, sampling $\gtrsim$3 phases for each star. We
obtained atmospheric parameters (T$_{\rm{eff}}$, log$g$, v$_{\rm{t}}$, and
[M/H]) and abundances of several iron-peak and $\alpha$-elements (Fe, Cr, Ni,
Mg, Ca, Si, and Ti) for different pulsational phases, obtaining
$\langle$[$\alpha$/Fe]$\rangle$=+0.31$\pm$0.19~dex over the entire sample
covering --2.2$<$[Fe/H]$<$--1.1~dex. We find that silicon is indeed extremely
sensitive to the phase, as reported by previous authors, and cannot be reliably
determined. Apart from this, metallicities and abundance ratios are consistently
determined, regardless of the phase, within 0.10--0.15~dex,
although caution should be used in the range $0\la\phi\la0.15$.
Our results agree with literature determinations for both variable and
non-variable field stars, obtained with very different methods, including low
and high-resolution spectroscopy. W~Vir and BL~Her stars, at
least in the sampled phases, appear indistinguishable from RRab from the
spectroscopic analysis point of view. Our large sample, covering all pulsation
phases, confirms that chemical abundances can be obtained for RR~Lyrae with the
classical EW-based technique and static model atmospheres, even rather close to
the shock phases.

\end{abstract}


\section{Introduction}

\begin{table*}
\caption{Basic information for the programme stars.}
\label{tab:basic}
\begin{tabular}{@{}lcccccllccc@{}}
\hline
Star     & R.A.(J2000) & Decl.(J2000)  & Type & alt (Type)       & $V$         & Epoch        & Period    & $A(V)$ & [Fe/H]$_{\mathrm{K06}}$ & [Fe/H]$_{\mathrm{B00}}$\\
         & (h m s) & ($\degr$ $\arcmin$ $\arcsec$) &           & & (mag)         & (JD 2400000$+$)  & (day)  & (mag)  & (dex)                   & (dex) \\
\hline
DR And*  & 01 05 10.71 & $+$34 13 06.3 & RRab & {\em (W Uma)}& 11.65 -- 12.94 & 51453.158583 & 0.5631300  & 1.10   & $-$1.42$\pm$0.20 & $-$1.48 \\
X Ari    & 03 08 30.88 & $+$10 26 45.2 & RRab &              & 11.28 -- 12.60 & 54107.2779   & 0.6511681  & 0.88   & $-$2.08$\pm$0.17 & $-$2.43 \\
TW Boo   & 14 45 05.94 & $+$41 01 44.1 & RRab &              & 10.63 -- 11.68 & 53918.4570   & 0.53226977 & 1.06   & $-$1.53$\pm$0.16 & $-$1.46 \\
TW Cap   & 20 14 28.42 & $-$13 50 07.9 & CWa  &              &  9.95 -- 11.28 & 51450.139016 & 28.610100  & (0.84) & ---              & $-$1.20 \\
RX Cet   & 00 33 38.28 & $-$15 29 14.9 & RRab &              & 11.01 -- 11.75 & 52172.1923   & 0.5736856  & 0.60   & $-$1.09$\pm$0.32 & $-$1.28 \\
U Com    & 12 40 03.20 & $+$27 29 56.1 & RRc  &              & 11.50 -- 11.97 & 51608.348633 & 0.2927382  & (0.34) & ---              & $-$1.25 \\
UZ CVn   & 12 30 27.70 & $+$40 30 31.9 & RRab &              & 11.30 -- 12.00 & 51549.365683 & 0.6977829  & 1.03   & $-$2.10$\pm$0.17 & $-$1.89 \\
AE Dra   & 18 27 06.63 & $+$55 29 32.8 & RRab &              & 12.40 -- 13.38 & 51336.369463 & 0.6026728  & 1.16   & $-$1.88$\pm$0.17 & $-$1.54 \\
BK Eri   & 02 49 55.88 & $-$01 25 12.9 & RRab &              & 12.00 -- 13.05 & 51462.198773 & 0.5481494  & 0.86   & ---              & $-$1.64 \\
UY Eri   & 03 13 39.13 & $-$10 26 32.4 & CWb  &              & 10.93 -- 11.66 & 51497.232193 & 2.2132350  & (0.64) & ---              & $-$1.60 \\
SZ Gem   & 07 53 43.45 & $+$19 16 23.9 & RRab &              & 10.98 -- 12.25 & 51600.336523 & 0.5011365  & 1.27   & $-$1.67$\pm$0.16 & $-$1.46 \\
VX Her   & 16 30 40.80 & $+$18 22 00.6 & RRab &              &  9.89 -- 11.21 & 53919.451    & 0.45536088 & 1.27   & $-$1.40$\pm$0.16 & $-$1.58 \\
DH Hya*  & 09 00 14.83 & $-$09 46 44.1 & RRab & {\em (W Uma)}& 11.36 -- 12.65 & 51526.426583 & 0.4889982  & 1.28   & $-$1.73$\pm$0.16 & $-$1.55 \\
V Ind    & 21 11 29.91 & $-$45 04 28.4 & RRab &              &  9.12 -- 10.48 & 47812.668    & 0.479601   & (1.06) & ---              & $-$1.50 \\
SS Leo   & 11 33 54.50 & $-$00 02 00.0 & RRab &              & 10.38 -- 11.56 & 53050.565    & 0.626335   & 1.06   & $-$1.93$\pm$0.17 & $-$1.79 \\
V716 Oph & 16 30 49.47 & $-$05 30 19.5 & CWb  &              &  8.97 --  9.95 & 51306.272953 & 1.1159157 & (1.39) & ---              & $-$1.55 \\
VW Scl   & 01 18 14.97 & $-$39 12 44.9 & RRab &              & 10.40 -- 11.40 & 27809.381    & 0.5109147 & (1.23) & ---              & $-$0.84 \\
BK Tuc   & 23 29 33.33 & $-$72 32 40.0 & RRab &              & 12.40 -- 13.30 & 36735.605    & 0.5502000 & (0.94) & ---              & $-$1.82 \\
TU UMa   & 11 29 48.49 & $+$30 04 02.4 & RRab &              &  9.26 -- 10.24 & 51629.148846 & 0.5576587 & 0.96   & $-$1.66$\pm$0.17 & $-$1.51 \\
RV UMa   & 13 33 18.09 & $+$53 59 14.6 & RRab &              &  9.81 -- 11.30 & 51335.380433 & 0.4680600 & 1.07   & $-$1.20$\pm$0.18 & $-$1.20 \\
UV Vir   & 12 21 16.74 & $+$00 22 03.0 & RRab &              & 11.35 -- 12.35 & 51579.459853 & 0.5870824 & 1.00   & $-$1.73$\pm$0.18 & $-$1.19 \\
\hline
\end{tabular}
\begin{flushleft} 
{\em Notes.} Coordinates (columns 2 and 3), type of variability (columns 4 and
5), rough magnitude ranges (column 6), and adopted periods (column 8) are from
the General Catalog of Variable Stars \citep[GCVS,][]{samus10}, except for
X~Ari, TW~Boo, RX~Cet, and VX~Her, that are known to have a variable period
\citep[][see also text]{leborgne07}. Stars marked with an asterisk do not have a
clear classification, but for our computations we used the type in column 4.
Epochs of maximum light (column 7) have been derived by us from ROTSE light
curves, except for X~Ari, TW~Boo, RX~Cet, and VX~Her, where a parabolic fit was
used, based on data from \citet{leborgne07}. For a few stars with no ROTSE data,
epochs are taken from the GCVS, and for SS~Leo from
\citet{maintz05}. $V$ amplitudes (column 9) are from \citet{kinemuchi06};
values in brackets were either derived from ROTSE light curves (TW~Cap, U~Com,
UY~Eri, and V716~Oph), or taken from the ASAS-3 catalogue  \citep[V~Ind, VW~Scl,
BK~Tuc;][]{asas,meyer06}. Metallicities are from \citet[][column
10]{kinemuchi06} and \citet[][column 11; typical errors are on the order
0.1--0.2~dex]{beers00}.
\end{flushleft}
\end{table*}

RR Lyrae stars are old, metal-poor, horizontal branch pulsating variables that 
are undergoing quiescent helium burning in their centres. With typical periods 
of 0.2--1.0 day and magnitude variations in the visual band of 0.3--1.6 mag,
they are called RRab,  RRc or RRd \citep{bailey02,jerzykiewicz77}, respectively,
depending on whether they are pulsating in the radial  fundamental mode, radial
first overtone, or both modes simultaneously. Once appropriate corrections are
made for evolutionary effects and for the fact  that the mean intrisic absolute
magnitudes are not constant \citep[actually, they  correlate with
metallicity:][and references  therein]{cacciari03}, RR Lyrae turn out to be
excellent distance indicators, probing  stellar populations in the Milky Way and
beyond: RRab stars are routinely used  to trace tidal tails and streams in the
Galactic halo, with important  advantages compared to main-sequence turnoff
stars, blue horizontal branch  stars and K giants
\citep[e.g.][]{vivas08,sesar10,sesar13,drake13} and to probe the structure of
external systems \citep[e.g.][]{moretti14}. Several nearby
RR~Lyrae and variable stars do indeed have parallaxes in the literature, either
from Hipparcos \citep{hog00,vanleeuwen07} or from dedicated studies
\citep{benedict11}. In the near future, the
Gaia\footnote{http://www.cosmos.esa.int/web/gaia} ESA mission will provide
parallaxes to the $\mu$as level for all nearby stars. Photometry, which has
been traditionally used for RR Lyrae distance determination, needs to be
complemented by spectroscopic information not only to account correctly for the
effect of metallicity on the absolute magnitude of these stars, but also to give
information on their kinematics and detailed chemical abundances (whenever
possible) which both help discriminate among various sub-populations with
different characteristics. 

However, while extensive literature exists concerning photometry and 
low-resolution spectroscopy of RR Lyrae stars, high-resolution spectroscopic 
studies are by far less numerous. This is most likely due to the limits on the 
exposure times, which translate into limits on the attainable S/N
(signal-to-noise) ratios in the spectra, imposed by the short pulsation periods.
Yet, a number of authors
\citep{butler76,butler79,clementini95,lambert96,preston06,kolenberg10,for11,kinman12,govea14}
did perform detailed chemical composition analyses based on high-resolution
spectra for different samples of RR Lyrae stars. The most important general
conclusions  that can be drawn from these studies are: (i) lines of most species
form in  local thermal equilibrium (LTE) conditions, thus standard LTE analyses
can be  performed, but see our discussions in
Sections~\ref{sec:grav} and \ref{sec:iron} for more details; (ii) co-addition of
spectra can be safely used to increase the S/N  ratios; (iii) the strongest and
more symmetric lines are found at phases 0.3$<  \phi <$0.5; (iv) effective
temperature (T$_{\rm{eff}}$), gravity ($\log g$), and microturbulent velocity
(v$_{\rm{t}}$) variations with phase  are regular, but abundance ratios are
mostly insensitive to phase.

High-resolution spectroscopic studies are mandatory in order to gain  insights
on the atmospheric behaviour of RR Lyrae stars. Our knowledge of  the
atmospheric dynamics in RR Lyrae stars is, in fact, still very poor. For 
instance, the physical origin of the Blazhko effect -- so named after Sergei 
Nikolaevich Blazhko, who was the first to report a long-period modulation of 
the lightcurve of a RR Lyrae star \citep{blazhko07} -- remains frustratingly 
elusive \citep[e.g.][and references therein]{chadid08,chadid13}, in spite of 
the fact that a significant fraction of RRab stars 
\citep[up to 50 per cent;][]{jurcsik09} exhibits these long-term modulations of 
amplitudes and phases. We refer the interested reader to \citet{fossati14} for
the most recent high-resolution spectroscopic study of RR~Lyr, dealing with all
the above issues.

Population~II Cepheids also play a relevant role as distance indicators and old
stellar populations tracers. They are generally classified as BL~Her, W~Vir and
RV~Tau according to their periods and evolutionary stages: BL~Her stars have the
shortest periods (0.8--4~days) and are evolving from the horizontal branch
towards the asymptotic giant branch (AGB); W~Vir stars, with periods in the
range 4--20 days, are crossing the instability strip during their blue-loop
excursions; and RV~Tau stars are in their post-AGB phase \citep[see
e.g.][]{soszynski08}. Despite some hints of unusual chemical compositions, they
have received scant attention from spectroscopists. A notable exception is the
study by \citet{maas07}, who analysed a sample of 19  BL~Her and W~Vir stars and
related the contrasting abundance anomalies to the different stars' evolutive
stages from the blue horizontal branch. Two of the three Population~II Cepheids
in our sample have abundance determinations from \citet{maas07}: TW~Cap and
UY~Eri.

\begin{table*}{}
\caption{Observing logs.}
\label{tab:logs}
\begin{tabular}{@{}lclcccrlclcccr@{}}
\hline
Star     & Inst. &Exp$_n$& Observation date   &Phase&t$_{\rm{exp}}$ & S/N&Star& Inst. &Exp$_n$ &Observation date&Phase&t$_{\rm{exp}}$ & S/N\\
         &       &     & (HJD 2400000+)         &      &(min)& &       & &     & (HJD 2400000+)         &      &(min)& \\
\hline
DR And   & SARG  &  1  & 55084.57606 & 0.63 & 30 &  80  & V Ind   & FEROS &  1* & 54302.85087 & 0.46 & 12 &  80 \\
         & SARG  &  2  & 55084.60982 & 0.69 & 60 &  80  &         & FEROS &  2* & 54302.92066 & 0.61 & 15 &  70 \\
         & SARG  &  3  & 55111.42640 & 0.31 & 60 &  80  &         & FEROS &  3* & 54302.93328 & 0.63 & 20 &  70 \\
X Ari    & APO   &  1* & 55521.74265 & 0.19 & 20 & 140  &         & HARPS &  1* & 54303.90886 & 0.67 & 20 &  30 \\
TW Boo   & SARG  &  1  & 54921.57757 & 0.61 & 30 &  70  &         & HARPS &  2* & 54303.91963 & 0.69 & 20 &  30 \\
         & SARG  &  2  & 54921.59904 & 0.65 & 30 &  70  &         & HARPS &  3* & 54303.93260 & 0.71 & 20 &  30 \\
         & SARG  &  3  & 54921.62052 & 0.69 & 30 &  70  &         & HARPS &  4* & 54304.72958 & 0.38 & 20 &  30 \\
TW Cap   & UVES  &  1  & 55070.65106 & 0.54 &  8 & 160  &         & HARPS &  5* & 54304.74426 & 0.41 & 20 &  30 \\
RX Cet   & SARG  &  1  & 55107.46135 & 0.51 & 45 & 130  &         & HARPS &  6* & 54304.90275 & 0.74 & 20 &  20 \\
U Com    & SARG  &  1  & 54922.40036 & 0.87 & 30 &  60  &         & HARPS &  7* & 54304.91729 & 0.77 & 20 &  20 \\
         & SARG  &  2  & 54922.42337 & 0.95 & 30 &  60  &         & HARPS &  8* & 54304.93168 & 0.80 & 20 &  20 \\
         & SARG  &  3  & 54922.44483 & 0.02 & 30 &  60  &         & HARPS &  9* & 54305.90708 & 0.83 & 30 &  20 \\
UZ CVn   & SARG  &  1  & 54922.47495 & 0.04 & 45 &  70  &         & HARPS &  10*& 54305.92491 & 0.87 & 30 &  30 \\
         & SARG  &  2  & 54922.50919 & 0.08 & 45 &  60  & SS Leo  & UVES  &  1  & 54951.53090 & 0.06 &7.5 &  65 \\
         & SARG  &  3  & 54922.54108 & 0.13 & 45 &  60  &         & UVES  &  2  & 54951.53683 & 0.07 &7.5 &  85 \\ 
AE Dra   & SARG  &  1  & 54921.70407 & 0.05 & 30 &  45  & V716 Oph& UVES  &  1  & 55081.50566 & 0.08 & 30 & 120 \\
         & SARG  &  2  & 54921.73314 & 0.10 & 45 &  50  & VW Scl  & UVES  &  1*  & 52168.62787 & 0.71 & 3 &  90 \\
BK Eri   & UVES  &  1* & 52169.88146 & 0.04 &  7 & 100  &         & UVES  &  2*  & 52167.87013 & 0.23 & 3 & 120 \\
         & UVES  &  2* & 52168.87630 & 0.20 &  7 &  80  &         & UVES  &  3*  & 52166.89349 & 0.32 & 3 &  90 \\
         & UVES  &  3* & 52167.74787 & 0.14 &  7 &  90  &         & UVES  &  4*  & 52165.78284 & 0.15 & 3 & 120 \\
         & UVES  &  4* & 52166.85011 & 0.51 &  7 &  70  &         & UVES  &  5*  & 52166.75294 & 0.05 & 3 & 150 \\
UY Eri   & SARG  &  1  & 55085.72750 & 0.38 & 45 & 130  & BK Tuc  & UVES  &  1  & 55050.75870 & 0.17 & 30 & 110 \\
SZ Gem   & SARG  &  1  & 55151.64242 & 0.50 & 45 &  80  &         & UVES  &  2  & 55050.80812 & 0.26 & 30 &  80 \\
VX Her   & SARG  &  1  & 54921.63477 & 0.86 & 25 &  70  &         & UVES  &  3  & 55051.84590 & 0.15 & 30 &  70 \\
         & SARG  &  2  & 54921.66660 & 0.05 & 30 &  70  & TU UMa  & SARG  &  1  & 54922.37438 & 0.45 & 30 & 110 \\
DH Hya   & UVES  &  1  & 54936.60042 & 0.79 & 13 &  70  & RV UMa  & SARG  &  1  & 54921.47819 & 0.62 & 30 &  60 \\
         & SARG  &  1  & 54921.40162 & 0.71 & 45 &  35  &         & SARG  &  2  & 54921.50074 & 0.66 & 30 &  80 \\
         & SARG  &  2  & 54921.43822 & 0.79 & 45 &  50  &         & SARG  &  3  & 54921.52265 & 0.71 & 30 &  80 \\
 V Ind   & UVES  &  1  & 55026.91751 & 0.19 &  4 & 120  & UV Vir  & UVES  &  1  & 54951.58967 & 0.87 & 20 & 120 \\
         & UVES  &  2  & 55069.64846 & 0.29 &  4 & 120  & \\
\hline
\end{tabular}
\begin{flushleft} 
\emph{Notes.} Archival spectra are marked with an asterisk in columns 3 and 10. The
listed S/N are evaluated @6000 \AA.
\end{flushleft}
\end{table*}

In this paper, we present atmospheric parameters, metallicities, and abundances 
of several iron-peak and $\alpha$-elements for 18 RR Lyrae
stars and 3 Population II Cepheids, observed at different phases. The paper is
organized as follows. In Section~\ref{sec:obs} we summarize the observations and
data reduction. In Section~\ref{sec:abuan} we describe the adopted model
atmospheres, atomic linelist and abundance analysis tool, and present the
equivalent width (EW) measurements. Section~\ref{sec:res} presents the results
of our spectroscopic abundance analysis, starting from the astrophysical
parameters determination and ending with the metallicities and abundance ratios
of our sample stars. Finally, the results are discussed --- and some conclusions
are drawn --- in Section~\ref{sec:disandcon}.

\section{Observations and data reduction}
\label{sec:obs}

The observations for the 18 RR Lyrae and 3 Population~II
Cepheids analysed in this paper have been obtained with two different telescopes
and instrumental set-ups and complemented with archive data. Basic literature
information for the programme stars can be found in Table~\ref{tab:basic} (see
also Section~\ref{sec:lit}). 

\subsection{SARG data}
\label{sec:sarg}

\begin{figure}
\centering
\includegraphics[width=\columnwidth]{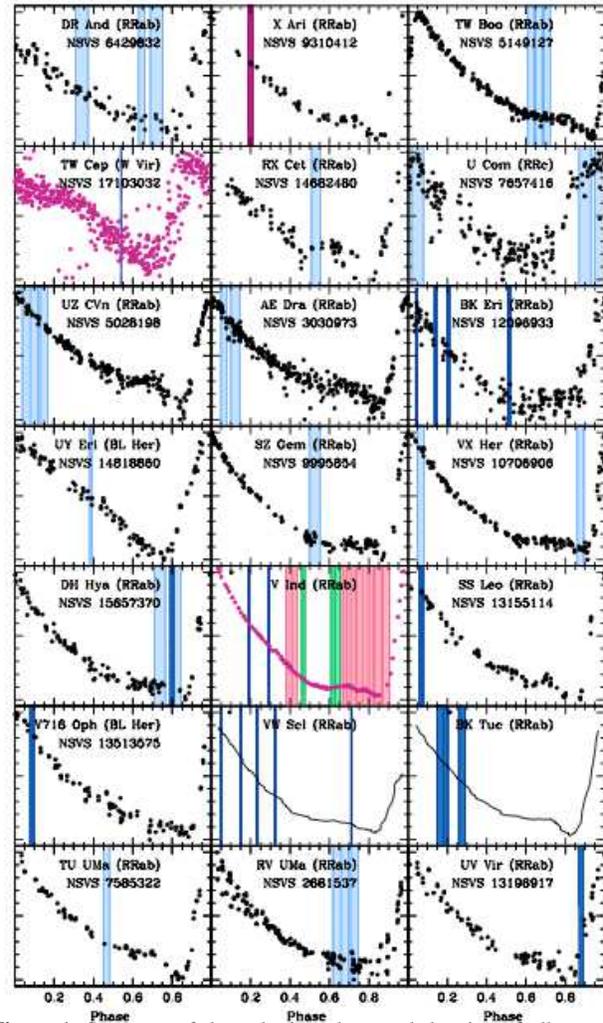}
\vspace{-1cm}
\caption[]{Summary of the pulsation phase and duration of all our exposures: 
SARG spectra are indicated by light blue stripes; UVES ones by dark blue
stripes; archival FEROS spectra by green stripes; archival HARPS spectra by
pink stripes; and the APO spectrum of X~Ari by a purple stripe. Each panel 
reports data for one star on an arbitrary scale. Whenever ROTSE light curve
data were available,  they were plotted as black dots on an arbitrary vertical
scale, just for  reference, and the ROTSE designation for the star was
indicated on each panel. For V~Ind, we used data from \citet[][purple
dots]{clementini90}; for TW~Cap we plotted the ASAS-3 data \citep[][purple
dots]{asas,meyer06}; for VW~Scl and BK~Tuc we obtained  template light curves
from data of stars with similar  characteristics:  SS~For \citep{cacciari87}
and TU~Uma \citep{liu89,fernley97}, respectively,  plotted as solid curves.}
\label{fig:logs}
\end{figure}

Observations of 15 RR Lyrae stars (DR~And, X~Ari, TW~Boo,
RZ~Cam, RX~Cet, U~Com, RV~CrB, SW~CVn, UZ~CVn, AE~Dra, SZ~Gem,
VX~Her, DH~Hya, TU~UMa and RV~UMa)\footnote{The quality of SARG
spectra of RZ~Cam, RV~CrB, SW~CVn, and X~Ari was not sufficient to obtain
reliable atmospheric parameters and abundances. Therefore, RZ~Cam, RV~CrB and
SW~CVn were not analysed, and will not be reported in the following, whereas
X~Ari was analyzed using archival spectra.} and one BL~Her star (UY~Eri) were
carried out with SARG@TNG \citep{sarg}, operated on the island of La Palma,
Spain, during two separate runs in March and between September and November
2009. During the first run (in visitor mode), variables were observed almost
always at random phases, while in the second run (in service mode) observations
were planned at minimum light. The observing conditions in both runs were
reasonably good, although non-photometric. Observations were generally split
into three exposures (see Table~\ref{tab:logs}); in some cases exposure times as
long as 45 or 60~min were necessary to gather enough S/N. This
implies that some of these spectra will suffer from significant line smearing,
where the distorted profiles from slightly different phases overlap, producing
additional distortions in the line shapes. The exact phase and phase coverage
of each exposure can be desumed from Table~\ref{tab:logs} or
Figure~\ref{fig:logs}. SARG was set-up to reach a resolving power of
R=$\delta\lambda/\lambda\simeq$30\,000 and to cover a spectral range from 4000
to 8500 \AA. The relatively low spectral resolution of SARG does
not damage significantly the already widened line profiles of pulsating
variables. We could reach a S/N roughly between 50 and 100 per pixel except on
the margins of the spectra, and we discarded spectra below S/N$\simeq$20--30,
approximately.

SARG spectra were reduced with the IRAF\footnote{IRAF (http://iraf.noao.edu/)
is distributed by the National Optical Astronomical Observatory, which is
operated by the Association of Universities for Research in Astronomy (AURA)
under cooperative agreement with the National Science Foundation.} tasks in the
{\em echelle} package. First of all we applied bias subtraction and
flat-fielding; for preparing flatfields (with {\em apflatten}) and also for
tracing spectra of faint stars we used spectra of the brightest star (one for
each night) to locate echelle orders. The position of each order was then
traced interactively with a cubic spline. Two-dimensional dispersion solutions
were found for the Th-Ar spectra; the typical r.m.s. deviation of the lines
from fitted wavelength calibration polynomial was near 0.03 \AA.  Sky
absorbtion lines (telluric bands of O$_{2}$ and H$_{2}$O) were removed using
the IRAF task {\em telluric} with the help of our own library of observed
spectra of fast rotating hot stars accumulated during the years. In particular,
the stars which provided best results in the case of SARG spectra were HR~5206
and HD~6215 observed with UVES in June 2000 at a slightly higher resolution.
Once the spectra were wavelength calibrated and extracted (with optimal
extraction), the orders were merged into a single spectrum by means of an S/N
weighted sum using the IRAF tasks {\em scombine} and {\em continuum}. An
example of the quality of the SARG spectra is shown in Figure~\ref{fig-spec}.

\subsection{UVES data}
\label{sec:uves}

Eight stars (SW~Aqr, TW~Cap, DH~Hya, V~Ind,
SS~Leo, V716~Oph, BK~Tuc and
UV~Vir)\footnote{The quality of the spectrum of SW~Aqr was not
sufficient for an abundance analysis, so SW~Aqr will not be reported in the
following.}  were observed with UVES@VLT \citep{uves,dekker00}, between April
and August 2009 in service mode. Observing conditions were clear, but mostly
non-photometric. Being the VLT more efficient, we needed generally shorter
exposures than with SARG to reach a similar S/N, resulting in less altered line
profiles, covering a shorter range of phases. The observing logs can be found in
Table~\ref{tab:logs} and the phase coverage can be seen in
Figure~\ref{fig:logs}. UVES was set-up to reach a resolution of R$\simeq$45\,000
and to cover a spectral range of 4500--7500~\AA; the S/N was slightly higher
than for SARG spectra, ranging between 70 and 150 per pixel, roughly.

The data reductions were performed with the UVES pipeline \citep{uvespipe} by
ESO as part of the service observations. The pipeline reductions include the
classical steps of bias subtraction, flat-field correction, wavelength
calibration and spectra extraction by means of optimal extraction, sky
subtraction, and finally order merging with pixel resampling. We normalized the
pipeline-processed spectra and corrected them for telluric absorption as done
for the SARG spectra.

\subsection{Archival data}
\label{sec:archive}

Additional spectra were retrieved from the ESO Advanced Data Products archive,
consisting of extracted and wavelength-calibrated spectra of BK~Eri (observed with
UVES), V~Ind (observed with FEROS and HARPS) and of VW~Scl (observed with UVES). A
spectrum of X Ari obtained with the ARC
Echelle Spectrograph (ARCES) at the Apache Point Observatory (S. Andrievsky, G.
Wallerstein, 2013, private communication) was also included in the sample. Information on these additional spectra
can be found in Table~\ref{tab:logs} and in Figure~\ref{fig:logs}. The spectra were normalised and corrected for telluric
absorption features following the procedures adopted for our own SARG and UVES
observations.

\begin{figure}
\centering
\includegraphics[width=\columnwidth]{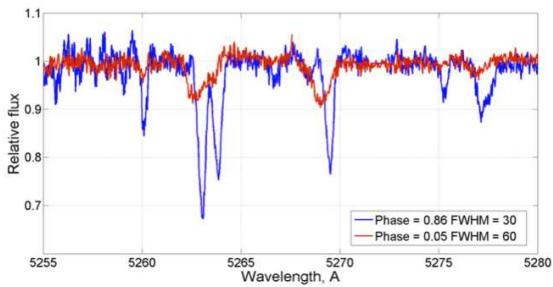}
\caption[]{Example of the quality of our SARG spectra. The two spectra of VX~Her
obtained with SARG are shown. According to \citet{kolenberg10} and as described
in the text, the $\phi$=0.05 spectrum should correspond to a shocked atmosphere
phase, while the $\phi$=0.86 spectrum should correspond to a quiescent phase. The
difference between the two spectra is striking \citep[see also figures 7, 8 and 9
of ][ for the effect of the shock on the hydrogen and metal lines of RR~Lyrae
stars in M4]{clementini94}.}
\label{fig-spec}
\end{figure}

\subsection{Reference literature information}
\label{sec:lit}

For all our programme candidates, we searched the literature for basic
information, which is listed in Tables~\ref{tab:basic} and \ref{tab:2mass}, and
displayed in Figure~\ref{fig:logs}. Epochs of maximum light were derived from
ROTSE light curves\footnote{http://www.rotse.net/} \citep[see also][for a
description of ROTSE variable star observation and analysis]{rotse} for most of
our targets, except for TW~Cap, V~Ind, VW~Scl, and BK~Tuc, for which epochs were
taken from the General Catalogue of Variable Stars \citep[GCVS,][]{samus10}, 
SS~Leo, for which an updated epoch was obtained from
\citet{maintz05}, and for three stars with varying period described below. The
periods and the Bailey variability types were obtained from the GCVS. Amplitudes
were obtained from \citet{kinemuchi06} for most stars, and when these were not
available, from ROTSE light curves by us or from the ASAS-3 catalogue \citep[All
Sky Automated Survey,][]{asas}. Also, 2MASS photometry and extinction data (see
Table~\ref{tab:2mass}) were used to derive photometric T$_{\rm{eff}}$ estimates
(Section~\ref{sec:teff}). Reference iron abundances were obtained from
\citet{beers00} and \citet[][see also Sections~\ref{sec:iron} and
\ref{sec:lit2}]{kinemuchi06}.

It is important to note that, according to \citet{fernley89},
X~Ari has varying period. More recently, \citet{leborgne07} found that also
TW~Boo, RX~Cet, and VX~Her have variable periods: using their reference data
(reported in Table~\ref{tab:basic}) and a parabolic fit (their equation~1), we
derived the epoch of maximum light closest to our observations, the appropriate
period, and finally the phase of each spectrum (reported in
Tables~\ref{tab:basic} and \ref{tab:logs}). 

\section{Abundance analysis}
\label{sec:abuan}

Many of our spectra were taken at random phase; Figure~\ref{fig:logs} summarizes
the phase and duration of our exposures: several of our spectra were taken away
from optimal phases. We discuss on the implications of using static atmosphere
models in Section~\ref{sec:static}, briefly reviewing theoretical and experimental
knowledge in the literature. Later (Sections~\ref{sec:ew} and \ref{sec:abu}) we
apply the classical method to all our spectra, regardless of their phase. Profiting
from the ample phase coverage, we then compare our results of both atmospheric
parameters and abundances \emph{a posteriori} (Section~\ref{sec:res}) across
different pulsation phases and with the literature, to assess the reliability
and repeatability of our analysis.

\begin{table*}
\caption{2MASS and extinction data used for determining the photometric temperatures listed in Table~\ref{tab:phototeff}.}
\label{tab:2mass}
\begin{tabular}{@{}lcccccccc@{}}
\hline
Star & 2MASS designation & 2MASS observation date & $\phi_{\mathrm{2MASS}}$ & $K_{\mathrm{2MASS}}$ & $A_V$ & $A_K$ & $\langle K \rangle$ \\ 
  & & (JD/BJD 2400000$+$) & (mag) & (mag) & (mag) & (mag) & (mag) \\
\hline
DR And   & J01051071$+$3413063 & 51105.8391/51105.84427 & 0.23 & 11.257 & 0.105 & 0.012 & 11.347 \\ 
X Ari    & J03083089$+$1026452 & 51519.7441/51519.74909 & 0.07 &  7.847 & 0.570 & 0.063 & 7.877 \\ 
TW Boo   & J14450595$+$4101442 & 50924.8937/50924.89713 & 0.79 & 10.378 & 0.041 & 0.004 & 10.248 \\ 
TW Cap   & J20142841$-$1350080 & 50968.9165/50968.92028 & 0.18 &  8.599 & 0.275 & 0.030 & --- \\ 
RX Cet   & J00333827$-$1529147 & 51399.8505/51399.85440 & 0.63 & 10.338 & 0.078 & 0.009 & 10.298 \\ 
U Com    & J12400319$+$2729561 & 51525.0186/51525.01801 & 0.34 & 10.993 & 0.043 & 0.005 & 11.003 \\ 
UZ CVn   & J12302770$+$4030320 & 51647.8253/51647.82894 & 0.10 & 10.815 & 0.065 & 0.007 & 10.875 \\ 
AE Dra   & J18270674$+$5529327 & 51630.9614/51630.96115 & 0.81 & 11.308 & 0.104 & 0.011 & 11.138 \\ 
BK Eri   & J02495585$-$0125118 & 51116.6763/51116.68175 & 0.66 & 11.568 & 0.149 & 0.016 & 11.528 \\ 
UY Eri   & J03133913$-$1026323 & 51105.7000/51105.70486 & 0.10 &  9.893 & 0.184 & 0.020 & --- \\ 
SZ Gem   & J07534345$+$1916240 & 50787.9584/50787.96258 & 0.93 & 10.845 & 0.121 & 0.013 & 10.745 \\ 
VX Her   & J16304079$+$1822005 & 51619.9286/51619.93043 & 0.22 &  9.508 & 0.133 & 0.015 & 9.588 \\ 
DH Hya   & J09001483$-$0946443 & 51197.6932/51197.69798 & 0.74 & 11.197 & 0.116 & 0.013 & 11.077 \\ 
V Ind    & J21112990$-$4504282 & 51403.7042/51403.70922 & 0.12 &  8.922 & 0.135 & 0.015 & --- \\ 
SS Leo   & J11335449-0002000   & 51198.7677/51198.77106 & 0.44 &  9.880 & 0.056 & 0.006 & 9.935 \\
V716 Oph & J16304946$-$0530195 & 51258.9032/51258.90542 & 0.55 &  9.862 & 1.202 & 0.132 & --- \\ 
VW Scl   & J01181495$-$3912448 & 51118.5535/51118.55669 & 0.43 & 10.066 & 0.046 & 0.005 & --- \\ 
BK Tuc   & J23293331$-$7232397 & 51528.5839/51528.58173 & 0.55 & 11.659 & 0.074 & 0.008 & --- \\ 
TU UMa   & J11294849$+$3004025 & 50882.8692/50882.87440 & 0.76 &  8.857 & 0.064 & 0.007 & 8.747 \\ 
RV UMa   & J13331810$+$5359146 & 51566.9156/51566.91780 & 0.67 &  9.816 & 0.053 & 0.006 & 9.776 \\ 
UV Vir   & J12211673$+$0022029 & 51599.7621/51599.76715 & 0.58 & 10.946 & 0.073 & 0.008 & 10.946 \\ 
\hline
\end{tabular}
\medskip
\begin{flushleft}
\emph{Notes.} 2MASS observation dates (JD) have been corrected to BJD using 
JSkyCalc (http://www.dartmouth.edu/~physics/faculty/skycalc/flyer.html). 
Extinction values (columns 6, 7) are from NED (http://ned.ipac.caltech.edu/) 
and are based on work by \citet{schlafly11}. $\langle K \rangle$ values have 
been calculated by us whenever possible from 2MASS data and $K$ light curve 
templates for RR~Lyrae by \citet{jones96}.\\
\end{flushleft}
\end{table*}

\subsection{Use of static model atmospheres}
\label{sec:static}

The anomalous features of hydrogen absorption lines (emission, line doubling) in
the spectra of RRab variables have been known for a long time
\citep{struve47,sanford49}, and were attributed to the existence of shock waves
at certain phases during the pulsation cycle, while there is no evidence to date
of shock waves in the atmosphere of RRc stars. Studying this phenomenon requires
spectroscopic observations with both high spectral and time resolution as well
as high S/N, and until quite recently it could only be done on a small number of
the nearest RRab variables with photographic spectra
\citep{preston64,prepac64,preston65,oke66}, and later with electronic light
detectors \citep{gillet88,gillet89}. The most accurate and detailed studies,
however, were obviously done during the last decade thanks to the use of more
advanced observation technology \citep{chadid08,kolenberg10,for11,preston11}.

Hydrodynamic model atmospheres \citep{hill72,fokin92,fokin99} identify two
pulsation phases where shocks occur: {\em (i)} the main shock corresponding to
the so-called `hump' at phase $\sim$0.9, when the infalling photosphere halts
and its outward acceleration rapidly increases; and {\em (ii)} the early shock
corresponding to the so-called `bump' at phase $\sim$0.7, likely produced by
colliding layers of material as the star approaches minimum radius
\citep{gillet88}. Fig.~\ref{fig-spec} shows two spectra of our sample star
VX~Her, one taken in a shocked phase, the other in a quiescent phase. Although
the shocks may fully develop in the higher atmospheric layers and hence mostly
affect hydrogen lines (see e.g. Oke et al. 1992), they are also detectable at
the photosphere, as shown by the broadening of photospheric FeI lines firstly
observed by \citet[][see also \citealt{clementini94}, and the quite long
discussion on shocks presented in that paper]{lebre93}. Both shocks are
associated with the emission of ultraviolet excess energy, which is stronger at
the 0.9 phase and weaker at the 0.7 phase, the origin of which has been
attributed to various physical causes \citep[see][for details]{smith95}.

\begin{figure}
\includegraphics[bb=200 70 596 765,angle=270,width=\columnwidth]{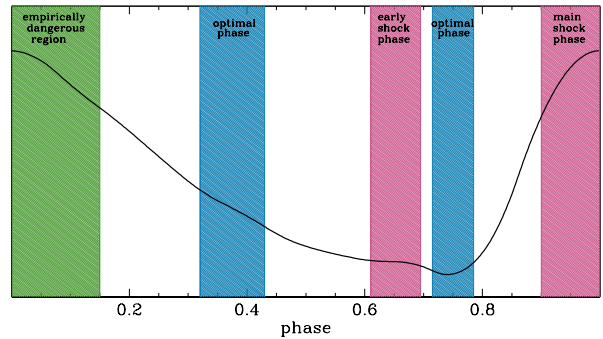}
\caption[]{A summary sketch of the quiescent (blue shaded regions) and shock
phases (red shaded regions) along the light-curve of a type ab RR~Lyrae (see
text for a discussion), on an arbitrary vertical scale. The
green region at $\phi<0.15$ is the empirically determined region where abundance
ratios appear to be poorly determined in some cases (see
Section~\ref{sec:ratios}).}
\label{fig:phase}
\end{figure}

Because static model atmospheres are more accurate and reliable than the
available hydrodynamic ones, the general approach has been so far to restrict
the analysis to the phase intervals in which the atmosphere is relatively
stable. Traditionally, this has been chosen around phase $\phi
\simeq$0.75--0.80 (see Figure~\ref{fig:phase}), corresponding to the minimum of
the typical light-curve of an ab-type RR Lyrae star, although the region can only be defined with a certain
level of approximation. More recently, \citet{kolenberg10} discussed the
possibility that $\phi \simeq$0.35 also corresponds to a phase of quiescence,
because RR~Lyrae stars reach their minimum radius and for a short time we can
safely use static model atmospheres (see Figure~\ref{fig:phase}). Their
statement is supported by the appearance of many iron lines with symmetric
shapes in spectra of RR~Lyrae observed in these phases, that can successfully
be used for a chemical analysis \citep{for11}.

From the empirical point of view, \citet{clementini95} performed abundance
analysis of RR~Lyrae spectra taken at or close to maximum light and \citet{for11}
and \citet{wallerstein12} compared the abundance analysis of stars observed at
different phases. Their conclusion is that reliable results can be obtained at
almost all phases, provided that one avoids the narrow regions around shocks (see
also Figure~\ref{fig:phase}) and, of course, that exposure times are as short as
possible to minimize the line shape deformations resulting from the overlap of
different phases. In the sample presented here, a few stars have
long exposure times (see Section~\ref{sec:sarg}) and a few exposures are into the
dangerous zones presented in Figure~\ref{fig:phase}. A discussion of these cases
is postponed to Section~\ref{sec:res}.

As mentioned above, RRc variables do not show evidence of having shock phases,
and the whole light curve is safe for abundance analysis \citep{govea14}. W Vir
variables, instead, do have shocks. \citet{maas07} reject spectra which show
line doubling, markedly asymmetric lines, or strong emission in the Balmer
lines. According to those authors, spectra not showing these characteristics are
likely to represent the atmosphere at a time when standard theoretical models
may be applied. However, as stressed by \citet{maas07}, this presumption should
be tested by analysis of a series of spectra taken over the pulsation cycle:
they obtained consistent results from the analysis of a limited number of stars
(three objects) with spectra taken at different phases. Among our programme
stars, TW~Cap is a W~Vir star, while UY~Eri and V716~Oph are BL~Her variables.
None of them exhibits strong profile alterations in their spectra, so we kept
them in our sample.

\subsection{Linelist and atomic data}
\label{sec:linelist}

To create a raw masterlist, we visually inspected the observed spectra, the
solar spectrum by \citet{moore66}, and a few synthetic spectra with temperatures
ranging from 5000 to 7000~K, gravities ranging from 2.5 to 3.5~dex, and
metallicities ranging from $-$2~dex to solar. The synthetic spectra were
computed with Tsymbal's \citep{tsymbal96} LTE code. All visible lines that
appeared not blended in at least one of the observed or theoretical spectra
(considering also molecular lines in the theoretical spectra) where identified
and included in the raw masterlist. 

Atomic data for the selected lines were obtained from the
VALD2\footnote{http://vald.astro.univie.ac.at/$\sim$vald/php/vald.php} and
VALD3\footnote{http://vald.astro.uu.se/} online databases \citep{vald}, including
line broadening parameters, when available. The employed oscillator strengths
(log$gf$) and excitation potentials ($\chi_{\rm{ex}}$) are reported in
Table~\ref{tab:ew}, along with the measured EWs for each spectrum. More in
detail, the major sources of log$gf$ data for the selected lines are: for Fe the
VALD2 critical compilation, based on 27 different sources\footnote{See
http://vald.astro.univie.ac.at/$\sim$vald/php/vald.php?docpage=datasets.txt for
more information.}; for Mg the Kurucz data (CD Rom 18)\footnote{See
http://kurucz.harvard.edu/atoms/ for the complete collection of Kurucz atomic
data.}; for Ca the Kurucz data (CD Rom 20-22); for Si the Kurucz data, from the
2007 set; for Ti~I the Kurucz data (CD Rom 18); for Ti~II  \citet{pickering01}
and \citet{wiese01}; for Cr~I and Cr~II the Kurucz data (CD Rom 20-22); finally,
for Ni both the Kurucz data (CD Rom 20-22) and the \cite{fuhr88} data.

\begin{table}{}
\caption{EWs and atomic data for individual program stars spectra.}
\label{tab:ew}
\begin{tabular}{l@{\hspace{0.25cm}}c@{\hspace{0.25cm}}c@{\hspace{0.2cm}}c@{\hspace{0.2cm}}c@{\hspace{0.25cm}}c@{\hspace{0.25cm}}c@{\hspace{0.25cm}}c@{\hspace{0.25cm}}c}
\hline
Star    & Spectrum  & $\lambda$    & Species & EW      & $\delta$EW & log$gf$  & $\chi_{\rm{ex}}$ \\
        &           &  (\AA)       &         & (m\AA)  &  (m\AA)    & (dex)    & (eV)             \\
\hline
DR And  & SARG1     &  4647.434    &  Fe I   & 58.20   &  7.11      & --1.351  & 2.949  \\
DR And  & SARG1     &  4678.846    &  Fe I   & 43.30   &  8.60      & --0.833  & 3.602  \\
DR And  & SARG1     &  4707.275    &  Fe I   & 40.70   &  5.00      & --1.080  & 3.241  \\
DR And  & SARG1     &  4736.773    &  Fe I   & 40.50   &  4.42      & --0.752  & 3.211  \\
DR And  & SARG1     &  4872.138    &  Fe I   & 89.20   &  4.96      & --0.567  & 2.882  \\
DR And  & SARG1     &  4924.770    &  Fe I   & 41.30   &  6.30      & --2.178  & 2.279  \\
\hline
\end{tabular}
\emph{Note.} Only a portion of this table is shown here for guidance regarding
its form and content. A machine-readable version of the full table is available
as supporting information with the online version of the paper.
\end{table}

After the raw masterlist was assembled, an additional line
selection was applied, based on empirical criteria. First, only lines that were
actually measured in at least three different spectra (see next section) were
retained and passed on through the abundance analysis. A further selection was
performed iteratively after the abundance analysis rejection procedures (see
Section~\ref{sec:abu}), resulting in a final clean linelist of 352 lines of 9
species.

\subsection{Equivalent widths}
\label{sec:ew}

EW were then measured with the help of DAOSPEC \citep{daospec},
run through the automated parameters optimizer DOOp \citep{doop}. In some cases
the line profiles were deformed. For the long SARG exposure spectra, double
peaks and asymmetric profiles were observed, owing to the integration along
phases that in some cases were far from the equilibrium state. In all those
cases, we forced DAOSPEC to adopt a radial velocity consistent with the deepest
peaks, and the dominant line substructures. Clearly, the EW measurements by
Gaussian fits were not optimal and this is reflected in the higher than usual
(considering the S/N ratio) errors on the measurements, but also on the large
uncertainties on the abundance results (see Section~\ref{sec:abu}). 

The way we used to measure EW has an impact on the resulting v$_{\rm{t}}$ of
spectra with deformed line profiles: for those spectra where a ``main"
component could be identified, centered, and fit by the code, the FWHM would be
relatively smaller, leading to a ``normal" v$_{\rm{t}}$, more similar to the
field stars with unperturbed atmosphere. For those spectra with long exposure
times, where the lines are also shallow and different atmospheric effects were
included in the line profiles, a ``global" Gaussian fit of all substructures
would lead to a larger FWHM, and higher than usual resulting v$_{\rm{t}}$. This
is indeed the case, as discussed further in Section~\ref{sec:micro} and shown
in Figure~\ref{fig:gravmicro}. However, the classical EW method implicitly
tends to compensate for these effects, and as a result the iron abundances are
relatively stable and compare well with the literature (see also
Sections~\ref{sec:iron} and \ref{sec:lit2}).

The measured EW with their errors \citep[as computed by DAOSPEC, see][for a
detailed description]{daospec} can be found in Table~\ref{tab:ew}, where only
lines surviving the described selection procedures are displayed.

\subsection{Abundance computations}
\label{sec:abu}

Abundance calculations were performed with GALA \citep{gala}, which
automatically finds the best atmospheric parameters and abundances, based on the
Kurucz suite of abundance calculation programs \citep{kurucz,klinux}; we used
the Atlas9 grid of atmospheric models computed by
\citet{atlas}\footnote{http://wwwuser.oat.ts.astro.it/castelli/}.
Briefly, GALA uses the classical method based on EW
measurements, which refines an initial T$_{\rm{eff}}$ estimate by e\-ra\-sing
any trend of iron abundance, A(Fe), with excitation potential; refines
$v_{\rm{t}}$ by imposing that weak and strong lines give the same A(Fe); refines
$\log g$ by minimizing the difference between A(FeI) and A(FeII); and finally,
the method checks that there is no residual trend of A(Fe) with wavelength.
Practically, as our spectral ranges included saturated telluric bands (after
6800~\AA) that were difficult to remove, we ended up cutting the noisiest
spectra after 5800~\AA \ (in the worst cases) or 6500~\AA \ (in the less bad
cases).

GALA automatically selects lines based on three criteria: {\em (i)} their strength;
{\em (ii)} their measurement error, provided by DAOSPEC in our case; and {\em
(iii)} their discrepancy from the average [Fe/H] of the other lines. In the last
step of our raw master line list refinement, we rejected all those lines that
survived GALA's rejections in less than three spectra. After removing those lines
from the master line list, we repeated both DAOSPEC (with its line selection) and
GALA a few times, obtaining the final clean linelist described in
Section~\ref{sec:linelist} that was actually used for the final EW measurements and
abundance analysis.

\begin{table}
\caption{Photometric T$_{\rm{eff}}$ estimates from infrared 2MASS photometry
(see Table~\ref{tab:2mass}) for a subset of programme stars.}
\label{tab:phototeff}
\centering
\begin{tabular}{@{}lccccc@{}}
\hline
Star & Spectrum & $\phi_{\mathrm{obs}}$ & $K_{\mathrm{obs}}$ & $V_{\mathrm{obs}}$ & $T_{\mathrm{eff}}$\\
\hline
DR And & SARG 1 & 0.63 & 11.380 & 12.795 & 6058 \\
       & SARG 2 & 0.69 & 11.405 & 12.813 & 6070 \\
       & SARG 3 & 0.31 & 11.264 & 12.543 & 6294 \\
X Ari  & APO 1  & 0.19 &  7.818 &  9.381 & 6607 \\
TW Boo & SARG 1 & 0.61 & 10.254 & 11.575 & 6129 \\
       & SARG 2 & 0.65 & 10.275 & 11.575 & 6166 \\
       & SARG 3 & 0.69 & 10.300 & 11.590 & 6183 \\
RX Cet & SARG 1 & 0.51 & 10.281 & 11.586 & 6186 \\
U Com  & SARG 1 & 0.87 & 11.013 & 11.323 & 8460 \\
       & SARG 2 & 0.95 & 10.843 & 11.269 & 8106 \\
       & SARG 3 & 0.02 & 10.843 & 11.269 & 8106 \\
UZ CVn & SARG 1 & 0.04 & 10.825 & 11.778 & 6955 \\
       & SARG 2 & 0.08 & 10.817 & 11.852 & 6775 \\
       & SARG 3 & 0.13 & 10.808 & 11.921 & 6611 \\
AE Dra & SARG 1 & 0.05 & 11.088 & 12.087 & 6899 \\
       & SARG 2 & 0.10 & 11.071 & 12.218 & 6588 \\
BK Eri & UVES 1 & 0.04 & 11.495 & 12.320 & 7361 \\
       & UVES 2 & 0.20 & 11.445 & 12.599 & 6628 \\
       & UVES 3 & 0.14 & 11.461 & 12.524 & 6816 \\
       & UVES 4 & 0.51 & 11.511 & 12.919 & 6152 \\
SZ Gem & SARG 1 & 0.50 & 10.712 & 12.079 & 6183 \\
VX Her & SARG 1 & 0.86 &  9.796 & 11.127 & 6245 \\
       & SARG 2 & 0.05 &  9.541 & 10.213 & 7673 \\
DH Hya & UVES 1 & 0.79 & 11.219 & 12.627 & 6107 \\
       & SARG 1 & 0.71 & 11.152 & 12.537 & 6147 \\
       & SARG 2 & 0.79 & 11.219 & 12.627 & 6107 \\
SS Leo & UVES 1 & 0.06 &  9.880 & 10.714 & 7163 \\
       & UVES 2 & 0.07 &  9.887 & 10.732 & 7138 \\
TU UMa & SARG 1 & 0.45 &  8.705 &  9.954 & 6302 \\
RV UMa & SARG 1 & 0.62 &  9.801 & 11.168 & 6050 \\
       & SARG 2 & 0.66 &  9.818 & 11.174 & 6068 \\
       & SARG 3 & 0.71 &  9.851 & 11.198 & 6083 \\
UV Vir & UVES 1 & 0.87 & 11.146 & 12.218 & 6666 \\
\hline
\end{tabular}
\end{table}

\section{Results}
\label{sec:res}

The results of our spectroscopic abundance analysis are reported in
Tables~\ref{tab:res} and \ref{tab:ratios} and discussed in the following sections,
starting from the astrophysical parameter determination and following with the
elements abundance ratios.

\subsection{Effective temperature}
\label{sec:teff}

\begin{figure}
\includegraphics[angle=270,width=\columnwidth]{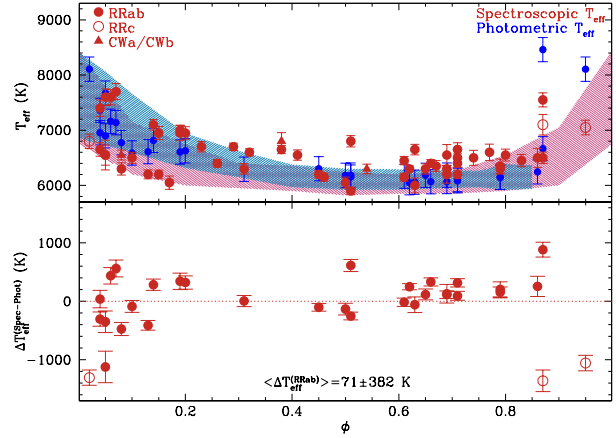}
\caption[]{\emph{Top panel:} comparison of our spectroscopic T$_{\rm{eff}}$ (red
symbols, where filled circles represent RRab, empty circles RRc, and filled
triangles Population~II Cepheids) with photometric ones. The blue circles are
derived from 2MASS colors (see text and Table~\ref{tab:phototeff}) while the
shaded areas represent different estimates from \cite{for11}: the blue shaded
region roughly covers the T$_{\rm{eff}}$ values of their eight comparison stars,
while the red shaded region roughly covers the T$_{\rm{eff}}$ values of their
programme stars (including uncertainties). \emph{Bottom panel:} difference (red
symbols) between the spectroscopic and the 2MASS photometric T$_{\rm{eff}}$
estimates.}
\label{fig:teff}
\end{figure}

To check that our spectroscopically derived effective temperatures are
reasonable, we compared with two different estimates of the expected
temperature at each phase, both based on photometry (see
Figure~\ref{fig:teff}).

The first method required an estimate of the infrared K magnitude at the phase
of each spectroscopic observation. We used the Two Micron All Sky Survey
\citep[2MASS;][]{2mass} data and K light-curve templates from \citet{jones96} to
obtain the K magnitudes of our programme stars at the phases of the
spectroscopic observations, $\phi_{\rm{obs}}$. We adopted the $V$ amplitudes
listed in Table~\ref{tab:basic}. We then derived the intrinsic $V-K$ color
corresponding to our $\phi_{\rm{obs}}$, by adopting extinction values from
\citet{schlafly11} and ROTSE light curves. A few stars with no ROTSE light
curves and Population~II Cepheids were excluded at this stage. Lastly, we used
the empirical calibration of T$_{\rm{eff}}$ versus colour by \citet[][their
equation~8]{alonso99}\footnote{Two stars have some spectra just outside the
limits of applicability of this calibration, having  (V--K)$<$0.1~mag: AE~Dra
(at $\phi$=0.05) and BK~Eri (at $\phi$=0.04 and 0.14). However, given the large
uncertainties involved in the procedure and our use of photometric temperatures
just as a reference value, we chose to use formula 8 by \citet{alonso99} in any
case.} to derive the T$_{\rm{eff}}$ values.

The T$_{\rm{eff}}$ values for the subset of our programme stars for which it was
possible to apply the method outlined above, are listed in
Table~\ref{tab:phototeff}. An estimate of the typical  uncertainty on the 2MASS
photometric T$_{\rm{eff}}$ was obtained by propagating  the reference magnitudes
uncertainties, and resulted of $\simeq$220~K. The  average difference between these
T$_{\rm{eff}}$ values and the corresponding  ones derived from spectroscopy is
$\langle \Delta$T$_{\rm{eff}} \rangle$=71$\pm$382~K.

\begin{table*}
\caption{Adopted atmospheric parameters and resulting iron abundances.}
\label{tab:res}
\begin{tabular}{lcccccccccccccc}
\hline
Star & Spectrum & T$_{\rm{eff}}$ & $\delta$T$_{\rm{eff}}$ & $\log g$ &
$\delta$log$g$ & v$_{\rm{t}}$ & $\delta$v$_{\rm{t}}$ & [FeI/H] &
$\sigma$[FeI/H] & $\partial$[FeI/H] & [FeII/H] & $\sigma$[FeII/H] &
$\partial$[FeII/H]\\
     &          & (K)            & (K)   & (dex)    & (dex) & (km/s)       
     & (km/s)& (dex)   & (dex) & (dex)    & (dex) & (dex)    & (dex)\\
\hline
DR And  & SARG 1  & 6000 & 134 & 1.60 & 0.10 & 2.7  & 0.2  & --1.40 & 0.19 & 0.07 & --1.45 & 0.03 & 0.03 \\
        & SARG 2  & 6200 & 160 & 2.30 & 0.17 & 2.9  & 0.4  & --1.53 & 0.17 & 0.07 & --1.61 & 0.07 & 0.03 \\
        & SARG 3  & 6300 &  95 & 2.10 & 0.14 & 2.4  & 0.2  & --1.29 & 0.12 & 0.07 & --1.33 & 0.13 & 0.03 \\
X Ari   & APO 2   & 6950 & 141 & 3.10 & 0.38 & 2.2  & 0.5  & --2.19 & 0.17 & 0.17 & --2.10 & 0.23 & 0.13 \\
TW Boo  & SARG 1  & 6150 &  71 & 2.00 & 0.09 & 2.1  & 0.2  & --1.49 & 0.12 & 0.07 & --1.45 & 0.11 & 0.04 \\
        & SARG 2  & 6300 &  93 & 2.30 & 0.06 & 2.2  & 0.2  & --1.39 & 0.14 & 0.07 & --1.37 & 0.03 & 0.04 \\
        & SARG 3  & 6300 &  64 & 2.10 & 0.07 & 2.6  & 0.3  & --1.49 & 0.11 & 0.07 & --1.49 & 0.33 & 0.03 \\
TW Cap  & UVES 1  & 6300 &  86 & 1.10 & 0.08 & 1.9  & 0.2  & --1.63 & 0.10 & 0.07 & --1.68 & 0.07 & 0.04 \\
RX Cet  & SARG 1  & 6800 & 103 & 2.00 & 0.12 & 1.7  & 0.2  & --1.38 & 0.13 & 0.07 & --1.39 & 0.09 & 0.04 \\
U Com   & SARG 1  & 7100 & 186 & 2.30 & 0.18 & 1.5  & 0.2  & --1.44 & 0.14 & 0.07 & --1.43 & 0.16 & 0.05 \\
        & SARG 2  & 7050 & 133 & 2.30 & 0.15 & 1.6  & 0.2  & --1.38 & 0.16 & 0.07 & --1.41 & 0.10 & 0.05 \\
        & SARG 3  & 6800 & 132 & 2.20 & 0.20 & 1.5  & 0.2  & --1.42 & 0.15 & 0.06 & --1.41 & 0.17 & 0.05 \\
UZ CVn  & SARG 1  & 6650 & 122 & 2.40 & 0.08 & 1.3  & 0.3  & --2.05 & 0.22 & 0.07 & --2.11 & 0.04 & 0.04 \\
        & SARG 2  & 6300 & 112 & 2.30 & 0.08 & 1.5  & 0.3  & --2.22 & 0.12 & 0.07 & --2.19 & 0.04 & 0.03 \\
        & SARG 3  & 6200 &  82 & 2.30 & 0.14 & 1.3  & 0.3  & --2.30 & 0.15 & 0.07 & --2.27 & 0.11 & 0.03 \\
AE Dra  & SARG 1  & 6550 & 184 & 1.90 & 0.11 & 1.6  & 0.2  & --1.51 & 0.18 & 0.07 & --1.45 & 0.10 & 0.04 \\
        & SARG 2  & 6500 & 102 & 1.80 & 0.19 & 1.6  & 0.2  & --1.44 & 0.15 & 0.07 & --1.42 & 0.12 & 0.04 \\
BK Eri  & UVES 1  & 7400 & 150 & 2.20 & 0.20 & 1.7  & 0.2  & --1.87 & 0.21 & 0.07 & --1.89 & 0.19 & 0.04 \\
        & UVES 2  & 6950 & 114 & 2.30 & 0.07 & 2.2  & 0.3  & --1.54 & 0.12 & 0.08 & --1.52 & 0.07 & 0.04 \\
        & UVES 3  & 7100 &  98 & 1.90 & 0.24 & 1.2  & 0.3  & --1.51 & 0.15 & 0.07 & --1.57 & 0.15 & 0.04 \\
        & UVES 4  & 5900 &  67 & 1.60 & 0.06 & 1.6  & 0.1  & --1.85 & 0.11 & 0.08 & --1.84 & 0.07 & 0.03 \\
UY Eri  & SARG 1  & 6800 & 157 & 1.80 & 0.08 & 1.6  & 0.4  & --1.43 & 0.19 & 0.07 & --1.48 & 0.07 & 0.04 \\
SZ Gem  & SARG 1  & 6050 & 101 & 1.90 & 0.10 & 1.5  & 0.1  & --1.65 & 0.13 & 0.07 & --1.59 & 0.10 & 0.04 \\
VX Her  & SARG 1  & 6500 & 172 & 1.90 & 0.09 & 2.3  & 0.2  & --1.49 & 0.10 & 0.07 & --1.51 & 0.05 & 0.04 \\
        & SARG 2  & 6550 & 268 & 2.20 & 0.11 & 2.3  & 1.1  & --1.73 & 0.19 & 0.06 & --1.76 & 0.03 & 0.04 \\
DH Hya  & UVES 1  & 6300 &  93 & 2.10 & 0.13 & 1.9  & 0.2  & --1.52 & 0.12 & 0.07 & --1.52 & 0.15 & 0.04 \\
        & SARG 2  & 6200 &  76 & 2.10 & 0.17 & 1.6  & 0.2  & --1.55 & 0.16 & 0.07 & --1.54 & 0.14 & 0.04 \\
        & SARG 3  & 6350 & 129 & 1.80 & 0.35 & 2.1  & 0.2  & --1.48 & 0.20 & 0.07 & --1.55 & 0.26 & 0.05 \\
V Ind   & UVES 1  & 7000 &  86 & 2.30 & 0.11 & 1.6  & 0.1  & --1.33 & 0.08 & 0.07 & --1.31 & 0.12 & 0.03 \\
        & UVES 2  & 6700 &  67 & 2.20 & 0.08 & 1.9  & 0.1  & --1.43 & 0.07 & 0.07 & --1.41 & 0.07 & 0.03 \\
        & FEROS 1 & 6150 &  72 & 1.70 & 0.05 & 1.8  & 0.1  & --1.56 & 0.09 & 0.07 & --1.59 & 0.05 & 0.04 \\
        & FEROS 2 & 6450 &  74 & 2.40 & 0.08 & 1.9  & 0.1  & --1.30 & 0.12 & 0.07 & --1.33 & 0.11 & 0.03 \\
        & FEROS 3 & 6650 &  95 & 2.60 & 0.10 & 1.9  & 0.2  & --1.20 & 0.23 & 0.06 & --1.19 & 0.04 & 0.04 \\
        & HARPS 1 & 6350 & 100 & 1.50 & 0.61 & 1.5  & 0.2  & --1.28 & 0.16 & 0.07 & --1.30 & 0.21 & 0.04 \\
        & HARPS 2 & 6550 & 206 & 2.30 & 0.93 & 2.8  & 0.3  & --1.17 & 0.10 & 0.06 & --1.18 & 0.30 & 0.04 \\
        & HARPS 3 & 6650 & 122 & 2.50 & 0.13 & 2.0  & 0.2  & --1.10 & 0.20 & 0.06 & --1.18 & 0.19 & 0.03 \\
        & HARPS 4 & 6650 &  75 & 2.50 & 0.08 & 1.4  & 0.1  & --1.25 & 0.09 & 0.06 & --1.24 & 0.13 & 0.03 \\
        & HARPS 5 & 6550 &  92 & 2.50 & 0.09 & 1.8  & 0.1  & --1.18 & 0.09 & 0.06 & --1.21 & 0.11 & 0.04 \\
        & HARPS 6 & 6500 & 135 & 2.50 & 0.17 & 2.8  & 0.3  & --1.30 & 0.27 & 0.07 & --1.37 & 0.04 & 0.04 \\
        & HARPS 7 & 6600 & 149 & 2.30 & 0.22 & 2.7  & 0.2  & --1.16 & 0.23 & 0.06 & --1.14 & 0.18 & 0.04 \\
        & HARPS 8 & 6550 & 108 & 2.60 & 0.18 & 2.7  & 0.3  & --1.23 & 0.25 & 0.06 & --1.25 & 0.14 & 0.04 \\
        & HARPS 9 & 6450 & 112 & 2.10 & 0.20 & 2.8  & 0.2  & --1.34 & 0.19 & 0.07 & --1.39 & 0.17 & 0.04 \\
        & HARPS 10& 6500 & 103 & 2.30 & 0.26 & 2.3  & 0.2  & --1.43 & 0.17 & 0.07 & --1.55 & 0.15 & 0.04 \\
SS Leo  & UVES 1  & 7600 & 140 & 2.50 & 0.07 & 1.1  & 1.1  & --1.49 & 0.08 & 0.18 & --1.47 & 0.05 & 0.13 \\
        & UVES 2  & 7700 & 145 & 2.50 & 0.07 & 1.6  & 1.0  & --1.46 & 0.07 & 0.19 & --1.46 & 0.06 & 0.16 \\
V716 Oph& UVES 1  & 6550 &  91 & 2.50 & 0.09 & 1.6  & 0.1  & --1.87 & 0.08 & 0.07 & --1.87 & 0.07 & 0.03 \\
VW Scl  & UVES 1  & 6400 &  76 & 2.50 & 0.09 & 2.4  & 0.2  & --1.18 & 0.11 & 0.07 & --1.26 & 0.11 & 0.03 \\
        & UVES 2  & 6700 &  94 & 2.10 & 0.08 & 1.9  & 0.1  & --1.32 & 0.08 & 0.07 & --1.31 & 0.06 & 0.04 \\
        & UVES 3  & 6600 &  64 & 2.30 & 0.11 & 1.7  & 0.1  & --1.16 & 0.10 & 0.06 & --1.23 & 0.12 & 0.04 \\
        & UVES 4  & 6950 & 118 & 2.30 & 0.17 & 1.7  & 0.2  & --1.30 & 0.10 & 0.07 & --1.30 & 0.11 & 0.04 \\
        & UVES 5  & 7600 & 150 & 2.30 & 0.20 & 1.6  & 0.2  & --1.46 & 0.15 & 0.09 & --1.43 & 0.10 & 0.04 \\
BK Tuc  & UVES 1  & 6050 & 120 & 1.90 & 0.11 & 2.4  & 0.1  & --1.80 & 0.10 & 0.08 & --1.76 & 0.06 & 0.04 \\
        & UVES 2  & 6400 &  68 & 2.30 & 0.10 & 2.3  & 0.4  & --1.60 & 0.11 & 0.07 & --1.63 & 0.08 & 0.05 \\
        & UVES 3  & 6200 &  79 & 2.00 & 0.08 & 2.1  & 0.1  & --1.62 & 0.12 & 0.07 & --1.65 & 0.08 & 0.04 \\
TU UMa  & SARG 1  & 6200 &  65 & 2.10 & 0.15 & 1.8  & 0.2  & --1.31 & 0.14 & 0.07 & --1.32 & 0.15 & 0.02 \\
RV UMa  & SARG 1  & 6300 &  53 & 2.20 & 0.10 & 1.6  & 0.1  & --1.27 & 0.12 & 0.06 & --1.23 & 0.13 & 0.04 \\
        & SARG 2  & 6400 &  69 & 2.30 & 0.12 & 1.8  & 0.1  & --1.16 & 0.10 & 0.07 & --1.16 & 0.14 & 0.05 \\
        & SARG 3  & 6400 &  71 & 2.30 & 0.12 & 1.9  & 0.1  & --1.13 & 0.13 & 0.06 & --1.20 & 0.12 & 0.03 \\
UV Vir  & UVES 1  & 7550 & 128 & 2.10 & 0.16 & 1.3  & 0.3  & --1.10 & 0.10 & 0.09 & --1.23 & 0.25 & 0.04 \\
\hline
\end{tabular}
\begin{flushleft}
\emph{Notes.} For each star and each spectrum, the columns report T$_{\rm{eff}}$,
log$g$, and v$_{\rm{t}}$ obtained with GALA with their formal errors, followed by
the average [FeI/H] and [FeII/H] obtained from  all Fe~I and Fe~II surviving lines,
their spreads $\sigma$[FeI/H] and $\sigma$[FeII/H], and their sensitivity 
$\partial$[FeI/H] and $\partial$[FeII/H] to variations of the atmospheric
parameters (see text for more details).
\end{flushleft}
\end{table*}

For a second comparison, we used the temperatures of the eight stars used by
\citet{for11} for creating their T$_{\rm{eff}}$-phase relations. Their
photometric temperatures (derived from B--V, V--R$_c$, and V--I$_c$ colors) are
used to derive a region of confidence, shown in Figure~\ref{fig:teff} as a blue
shaded area. We also used the final T$_{\rm{eff}}$ derived by \citet{for11} for
their programme stars (including uncertainties), to derive an additional region
of confidence (red shaded area in Figure~\ref{fig:teff}). The two regions
together cover a similar parameter space as that covered by our targets, both in
metallicity and in period. Our spectroscopically derived T$_{\rm{eff}}$ values
mostly fall inside or near the borders of the shaded areas.

We note that the spectrum of UV~Vir, taken dangerously close to the main shock
zone, needed a higher than expected T$_{\rm{eff}}$ to converge, but gravity and
microturbulence still appear reasonable. A special discussion deserves the case
of U~Com, the only type RRc variable, which displays a much lower spectroscopic
temperature (by more than 1000~K) than the photometric estimates, being more in
line with the temperatures expected for other RRab variables in the sample. The
only other high-resolution studies of RRc variables, to our knowledge, are:
\citet[][containing DH~Peg and T~Sex]{lambert96} and \citet[][who specifically
targeted eight RRc stars]{govea14}. Neither of these studies, covering
alltogether a lager metallicity range than ours, report T$_{\rm{eff}}>$7600~K; in
particular, Figure~11 by \citet{govea14} illustrates the variation of their
spectroscopic T$_{\rm{eff}}$ as a function of phase: for the phases of our three
spectra of U~Com, we should expect 7000$<$T$_{\rm{eff}}<$7500~K, roughly. We are
thus confident that our T$_{\rm{eff}}$ for this star are roughly correct. As
supporting evidence, we note that the resulting [Fe/H] for U~Com is only 0.2~dex
lower than the \citet{beers00} estimate. 

In conclusion, our spectroscopically derived T$_{\rm{eff}}$ values agree with the
values that are, roughly speaking, expected judging from optical \citep{for11}
and infrared (2MASS) photometry, except for a marginal discrepancy for some stars
around 0.3$<\phi<$0.4 and 0.6$<\phi<$0.8. The T$_{\rm{eff}}$ obtained for each
spectrum, along with the error estimated by GALA \citep[see][for more
details]{gala} from the slope of [Fe/H] versus excitation potential, are listed
in Table~\ref{tab:res}.

\subsection{Surface gravity}
\label{sec:grav}

\begin{figure}
\includegraphics[angle=270,width=\columnwidth]{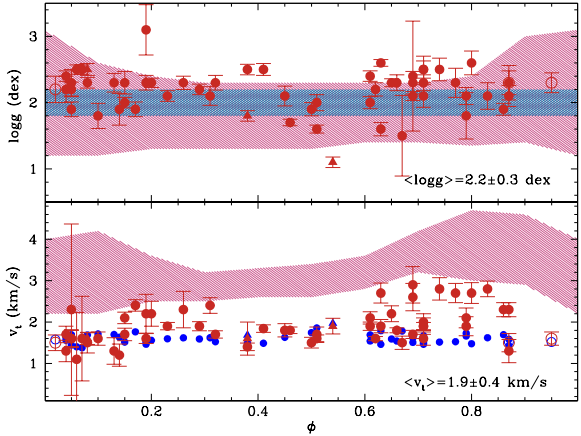}
\caption[]{Comparison of our spectroscopic $\log g$ (red symbols, upper panel)
and $v_{\rm{t}}$ (red symbols, bottom panel) with different literature reference
values. The red shaded regions represent roughly the final values obtained by
\citet{for11}, including their uncertainties. The blue shaded strip in the upper
panel shows the theoretical (fixed) $\log g$ adopted by \citet{for11} to build
their synthetic spectra. The blue filled circles in the lower panel show
v$_{\rm{t}}$ values obtained for our programme stars using the GES (Gaia-ESO
Survey) relation (see text) for non-variable stars.}
\label{fig:gravmicro}
\end{figure}

The acceleration term that is needed to account for the dynamic atmosphere of
RR~Lyrae \citep{clementini05} can be determined by differentiating the radial
velocity curve, which is a basic ingredient in the Baade-Wesselink
method\footnote{We found that a few of our programme stars were
previously analysed with the Baade-Wesselink method, based on very accurate
visual and infrared light curves; these are X~Ari \citep{fernley89}, V~Ind
\citep{clementini90} and TU~UMa \citep{liu90}.}. The effect of the early and
main shocks can be computed \citep[see for example the computations for S~Arae
by][their Figure~3]{chadid08}. In general, the acceleration associated with the
main shock should produce a significant increase of the effective gravity with
respect to the static value, $\Delta \log g \simeq$ 0.6--0.8 dex, whereas the
acceleration associated with the early shock should affect the gravity only
marginally, by $\le$0.1 dex. Apart from the absolute values of the gravity,
which depend on the assumed stellar mass and radius, the effective
gravity/acceleration curves show that the only part of the pulsation cycle where
gravity can be significantly different from the static value is around the
maximum light (approximately minimum radius) phase. Recent model computations by
\citet{kolenberg10} and observed spectra analysis by \citet{for11} show that
assuming a constant gravity throughout the pulsation cycle is appropriate
(within 0.1~dex). 

We used the spectroscopic method to balance Fe~I and Fe~II and our derived
gravities distribute flatly around $\langle \log g \rangle$= 2.2$\pm$0.3~dex
(see Table~\ref{tab:res} and Figure~\ref{fig:gravmicro}) and show no significant
trend with phase. In particular, we note that four of our spectra are taken in
the shocked zone with 0.85$<\phi<$1.0\footnote{They are: the second and third
SARG spectra of U~Com, the second SARG spectrum of VX~Her, and the first UVES
spectrum of V~Ind.}, but none displays largely deviant gravities. A large
scatter is anyway present, undoubtedly caused by the paucity of very reliable
Fe~II lines away from the optimal phases: X~Ari displays a very high
log$g$=3.1$\pm$0.38~dex, for that reason. 

Our log$g$ values are broadly compatible with the ones found by \citet{for11},
and are substantially lower than those obtained in past high-resolution studies
\citep{butler79,clementini95,lambert96}, which were closer to 3~dex and in
general to the B-W determinations of gravity. Explanations for this difference
were searched by \citet{for11}, who invoked uncertainties on the NLTE corrections
\citep[see also][]{clementini95,lambert96}: NLTE effects should in principle
produce lower gravities when one obtains gravity by forcing Fe~I and Fe~II to be
as close as possible\footnote{This is because LTE abundances of Fe~I should be
lower than those of Fe~II, and to compensate that, a lower gravity would be
needed in an LTE analysis \citep[see also][]{allende99}.}. While further
investigation of NLTE effects in the atmospheres of RR~Lyrae stars would be
highly desirable, past studies \citep[see for example][]{lambert96} do support
the idea that a 0.5--1.0~dex difference in log$g$ could be induced by NLTE
effects. Because the LTE spectroscopic log$g$ values could be in
principle indicative of NLTE effects, the fact that we do not observe significant
log$g$ changes with phase (nor with [Fe/H]), is suggestive of a relatively small
NLTE effect, contained within roughly 0.2~dex in terms of Fe~I, unless other
effects occurring in the complex atmospheres of these stars act to mask NLTE
effects: for example, the large discrepancy in the behaviour of log$g$ along the
pulsation cycle found with the B-W method (see above) and with high-resolution
spectroscopic analysis, is clearly not understood yet.

We also studied the impact of adopting different Fe~II log$gf$ (Fe~I atomic data
are overall more reliable), that are a well known source of
uncertainty in this type of analysis \citep{melendez09}. We measured an average
difference between the present study and the one by \citet{clementini95} --- as
an example --- of $\langle \log gf
\rangle$=+0.06$\pm$0.18~dex\footnote{Our log$gf$ scale for Fe~II
is 0.05$\pm$0.08~dex lower, on average, than that of \citet{melendez09}.}. We
thus tried changing our log$gf$ on our highest S/N spectrum (TW~Cap) by that
amount and allowed for GALA to re-converge with the new Fe~II log$gf$ values and
the same EWs. As a result, log$g$ was raised by only 0.2~dex, with
T$_{\rm{eff}}$ untouched and v$_{\rm{t}}$ raised by 0.2~km/s. While these
changes go in the right direction to reconcile the mentioned studies, they are
largely insufficient to explain the said differences.

To preserve the internal consistency of our analysis, and supported by the
findings by \citet{for11}, we used the log$g$ spectroscopic values obtained by
enforcing Fe ionization equilibrium, which are listed in Table~\ref{tab:res},
along with the error estimated by GALA \citep[see][for more details]{gala} from
the difference between [FeI/H] and [FeII/H].

\subsection{Microturbulent velocity}
\label{sec:micro}

As discussed in Section~\ref{sec:ew}, the method we use for measuring EWs has an
impact on our resulting v$_{\rm{t}}$ values obtained by balancing [FeI/H] with
EW. There are three cases: {\em(i)} the line profiles are not distorted and the
Gaussian fit is a good approximation; {\em(ii)} the line profiles are mildly
asymmetric when the atmosphere is not static, thus DAOSPEC could still fit
Gaussians with a slightly higher FWHM to include the whole line structure,
obtaining a reasonably reliable EW; {\em (iii)} the line profiles are heavily
distorted, for example because of a long exposure time, including different
phases where the line structures change rapidly, and DAOSPEC was forced to fit
the ``main" component of the line with a Gaussian, neglecting or leaving out
secondary components or asymmetric wings: the adopted FWHM is not much higher
than what expected for a non-variable star with similar
characteristics\footnote{It is interesting to note here that the expected axis
rotation of RRab stars does not generally exceed $v_{\rm{rot}}$sin$i$=6~km/s
\citep[see][and references therein]{preston13}, thus it is not surprising that
isolating the ``main" Gaussian component of a multiple or asymmetric line
profile produces FWHM values not much higher than the ones expected from the
spectral resolution.}. It is important to recall at this point that the method
employed by \citet{for11} was to integrate the heavily distorted lines instead
of fitting them with a Gaussian: this explains their v$_{\rm{t}}$ values, higher
than those of non-variable field stars with similar characteristics.

As can be seen in Figure~\ref{fig:gravmicro}, our resulting v$_{\rm{t}}$ values
are mostly similar to the ones expected for non-pulsating field stars, on the
basis of the Gaia-ESO survey v$_{\rm{t}}$ relation (Bergemann et al, in
preparation), and sometimes higher, almost as high as those by \citet{for11}.
The low values correspond to cases {\em (i)} and {\em (iii)} listed above, while
the intermediate or high values correspond to case {\em (ii)} above. We rarely
reach as high values as \citet{for11} or predicted by \citet{fokin99}, because
we rarely use the whole line profile for our abundance analysis. It is important
to note, however, that both methods for measuring EW produce reliable iron
abundances, because with both the adopted abundance analysis method ensures
self-consistency (though parameters inter-dependence) in this respect. The large
scatter of parameters in our case does, however, imply a larger uncertainty  for
these variable stars compared with non-variable stars observed with similar 
spectral quality (see below).

The v$_{\rm{t}}$ obtained for each spectrum, along with the error estimated by GALA
\citep[see][for more details]{gala} from the slope of the relation between [Fe/H]
and EW, are listed in Table~\ref{tab:res}.

\subsection{Iron abundance}
\label{sec:iron}

\begin{figure}
\includegraphics[angle=270,width=\columnwidth]{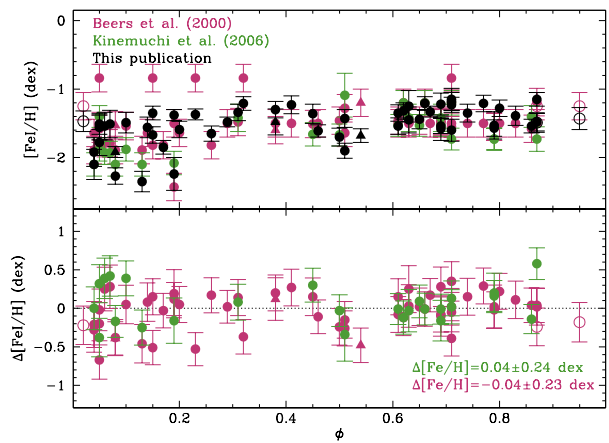}
\caption[]{Comparison of our spectroscopic [FeI/H], plotted in black, with
\citet{beers00}, plotted in magenta, and \citet{kinemuchi06}, plotted in green.
The symbols are the same as in Figs~\ref{fig:teff} and \ref{fig:gravmicro}. The
top panel shows the absolute values and the bottom panel the differences, in
the sense of our measurements minus the literature ones, as a function of
phase.}
\label{fig:met}
\end{figure}

We derived our Fe~I and Fe~II abundances as the median of all the available iron
lines, generally of the order of $\simeq$70--90 Fe~I and $\simeq$10 Fe~II
surviving lines. As we generally achieved a good ionization balance between Fe~I
and Fe~II ($\Delta$[Fe/H] was typically 0.01$\pm$0.04~dex, ranging from -0.09 to
+0.13~dex), we used Fe~I to define [Fe/H]. We do not detect very large non-LTE
effects, that with our method would appear as very low gravities, but according
to \citet{lambert96} and \citet{clementini95} these should be either of the
order of 0.2~dex or negligible, respectively, and thus could imply some
$\simeq$0.5--1.0~dex underestimate of the surface gravities (see
also the discussion in Section~\ref{sec:grav}). 

We compared our results with the two studies by \cite{kinemuchi06} and
\citet{beers00}. The former study used both the \citet{jurcsik96} and  the
\citet{sandage04} calibrations of RR~Lyrae metallicities as a function of the
period and the Fourier $\phi_{31}$ parameter. The latter study was instead based
on the \citet{beers99} recalibration of their prism survey to search for
metal-poor stars using the calcium K line. Both literature calibrations are thus
based on solar abundance values around A(Fe)$_{\odot}$=7.55--7.50 and thus
should agree well with our adopted solar composition \citep{solar}. For a more
detailed comparison with high-resolution spectroscopic studies, see
Section~\ref{sec:lit2}.

As shown in Figure~\ref{fig:met}, this is indeed the case: our values show no
significant offset with the two cited studies, and no significant trends with
phase are observed\footnote{A small oscillation can be picked up by the eye,
with an amplitude of $\lesssim$0.1~dex, almost hidden within the scatter. While
the oscillation is not statistically significant with the data in hand, it still
suggests that the use of spectra at different phases is risky when an error
below 0.1~dex is needed.}. The difference with \citet{beers00} is
$\langle\Delta$[Fe/H]$\rangle=$--0.04$\pm$0.23~dex, in the sense that our [Fe/H]
values are slightly lower than theirs; the difference with \citet{kinemuchi06}
is $\langle\Delta$[Fe/H]$\rangle=+$0.04$\pm$0.24~dex. \citet{beers00} report an
error on their estimates of 0.1--0.2~dex, while \citet{kinemuchi06} reported a
scatter of the order of 0.3~dex. The comparison shows that our global errors on
[Fe/H] of each single spectrum should be --- roughly speaking --- 0.10--0.15~dex
at most. Given the variety of [Fe/H] calibrations for RR~Lyrae in the
literature, all having reported uncertainties roughly around 0.2--0.3~dex
\citep[for example][to name a
few]{catelan92,sandage93,jurcsik96,alcock00,sandage04,bono07} we can conclude
that the agreement found in the metallicity range covered in this paper is more
than satisfactory. Also, being the present work one of the few based on high
resolution spectroscopy (R$\geq$30\,000, see also references in
Section~\ref{sec:lit2}), this lends independent support to those calibrations,
at least in the explored metallicity range.

The average Fe~I and Fe~II abundances of the surviving lines in each spectrum can
be found in Table~\ref{tab:res}, together with their spreads ($\sigma$[FeI/H] and
$\sigma$[FeII/H]) and their sensitivity to a variation of $\pm$100~K in
T$_{\rm{eff}}$, $\pm$0.1~dex in log$g$, and $\pm$0.1 km/s in v$_{\rm{t}}$
($\partial$[FeI/H] and $\partial$[FeII/H], computed by summing in quadrature the
abundance variations obtained by varying each parameter separately). As can be
seen, Fe~I is generally twice as sensitive to parameters variations than Fe~II.

The final Fe abundance for each star was computed as the weighted average of the
Fe~I measurements for each epoch, using as weights $w=1/\delta$[FeI/H]$^2$, where
$\delta$[FeI/H] was computed by summing in quadrature the random error
$\sigma$[FeI/H]$/\sqrt(n)$ \footnote{$n$ is the number of surviving Fe~I lines in
each spectrum.} and the variations of [FeI/H] obtained by altering the atmospheric
parameters by their errors, as listed in Table~\ref{tab:res}. The result is shown
in Table~\ref{tab:ratios}.

\begin{table*}
\caption{Abundance ratios.}
\label{tab:ratios}
\begin{tabular}{lrrrrrrrrrrr} 
\hline
Star     & [Fe/H]            & [Mg/Fe]         & [Ca/Fe]         & [Si/Fe]         & [Ti/Fe]         & [Cr/Fe]           & [Ni/Fe] \\
         & (dex)             & (dex)           & (dex)           & (dex)           & (dex)           & (dex)             & (dex) \\ 
\hline
DR And   & --1.37 $\pm$ 0.12 & 0.39 $\pm$ 0.41 & 0.24 $\pm$ 0.19 & 0.67 $\pm$ 0.21 & 0.20 $\pm$ 0.14 & --0.11 $\pm$ 0.26 &   0.09 $\pm$ 0.20 \\
X Ari    & --2.19 $\pm$ 0.17 & 0.50 $\pm$ 0.08 & 0.27 $\pm$ 0.11 & ...             & 0.73 $\pm$ 0.62 & --0.10 $\pm$ 0.40 &   0.30 $\pm$ 0.10 \\
TW Boo   & --1.47 $\pm$ 0.05 & 0.24 $\pm$ 0.10 & 0.24 $\pm$ 0.07 & 0.69 $\pm$ 0.02 & 0.20 $\pm$ 0.13 &   0.03 $\pm$ 0.17 &   0.07 $\pm$ 0.04 \\
TW Cap   & --1.63 $\pm$ 0.06 & 0.75 $\pm$ 0.07 & 0.28 $\pm$ 0.06 & 0.82 $\pm$ 0.19 & 0.20 $\pm$ 0.07 & --0.02 $\pm$ 0.13 & --0.04 $\pm$ 0.06 \\
RX Cet   & --1.38 $\pm$ 0.07 & 0.60 $\pm$ 0.12 & 0.18 $\pm$ 0.07 & 0.80 $\pm$ 0.18 & 0.11 $\pm$ 0.23 & --0.02 $\pm$ 0.08 & --0.04 $\pm$ 0.07 \\
U Com    & --1.41 $\pm$ 0.03 & 0.33 $\pm$ 0.19 & 0.23 $\pm$ 0.05 & 0.89 $\pm$ 0.35 & 0.18 $\pm$ 0.11 &   0.01 $\pm$ 0.09 &   0.05 $\pm$ 0.04 \\
UZ CVn   & --2.21 $\pm$ 0.13 & 0.48 $\pm$ 0.08 & 0.50 $\pm$ 0.24 & 1.43 $\pm$ 0.12 & 0.38 $\pm$ 0.18 &   0.09 $\pm$ 0.34 &   0.36 $\pm$ 0.52 \\
AE Dra   & --1.46 $\pm$ 0.05 & 0.46 $\pm$ 0.05 & 0.19 $\pm$ 0.04 & 1.08 $\pm$ 0.01 & 0.19 $\pm$ 0.19 &   0.28 $\pm$ 0.22 &   0.18 $\pm$ 0.16 \\
BK Eri   & --1.72 $\pm$ 0.21 & 0.46 $\pm$ 0.14 & 0.23 $\pm$ 0.15 & 0.52 $\pm$ 0.20 & 0.12 $\pm$ 0.27 & --0.04 $\pm$ 0.19 & --0.05 $\pm$ 0.34 \\
UY Eri   & --1.43 $\pm$ 0.11 & 0.41 $\pm$ 0.08 & 0.20 $\pm$ 0.11 & 0.42 $\pm$ 0.24 & 0.17 $\pm$ 0.27 & --0.01 $\pm$ 0.11 &   0.05 $\pm$ 0.13 \\
SZ Gem   & --1.65 $\pm$ 0.07 & 0.51 $\pm$ 0.08 & 0.34 $\pm$ 0.07 & 0.61 $\pm$ 0.10 & 0.12 $\pm$ 0.25 & --0.02 $\pm$ 0.29 & --0.11 $\pm$ 0.10 \\
VX Her   & --1.56 $\pm$ 0.17 & 0.17 $\pm$ 0.22 & 0.18 $\pm$ 0.25 & 1.17 $\pm$ 0.45 & 0.01 $\pm$ 0.22 &   0.08 $\pm$ 0.28 &   0.15 $\pm$ 0.05 \\
DH Hya   & --1.53 $\pm$ 0.01 & 0.38 $\pm$ 0.10 & 0.31 $\pm$ 0.04 & 0.63 $\pm$ 0.31 & 0.13 $\pm$ 0.20 &   0.03 $\pm$ 0.17 &   0.16 $\pm$ 0.05 \\
V Ind    & --1.30 $\pm$ 0.14 & 0.41 $\pm$ 0.12 & 0.23 $\pm$ 0.11 & 0.46 $\pm$ 0.29 & 0.15 $\pm$ 0.19 & --0.06 $\pm$ 0.14 & --0.10 $\pm$ 0.15 \\
SS Leo   & --1.48 $\pm$ 0.07 & 0.41 $\pm$ 0.13 & 0.35 $\pm$ 0.10 & ...             & 0.20 $\pm$ 0.06 & --0.05 $\pm$ 0.05 &   0.87 $\pm$ 0.21 \\
V716 Oph & --1.87 $\pm$ 0.06 & 0.51 $\pm$ 0.04 & 0.34 $\pm$ 0.05 & 0.59 $\pm$ 0.05 & 0.35 $\pm$ 0.13 &   0.01 $\pm$ 0.04 &   0.06 $\pm$ 0.10 \\
VW Scl   & --1.22 $\pm$ 0.10 & 0.38 $\pm$ 0.10 & 0.22 $\pm$ 0.12 & 0.47 $\pm$ 0.06 & 0.23 $\pm$ 0.30 & --0.08 $\pm$ 0.12 & --0.10 $\pm$ 0.07 \\
BK Tuc   & --1.65 $\pm$ 0.09 & 0.57 $\pm$ 0.11 & 0.32 $\pm$ 0.09 & 0.38 $\pm$ 0.06 & 0.13 $\pm$ 0.12 & --0.11 $\pm$ 0.09 &   0.01 $\pm$ 0.08 \\
TU UMa   & --1.31 $\pm$ 0.05 & 0.29 $\pm$ 0.03 & 0.25 $\pm$ 0.04 & 0.28 $\pm$ 0.05 & 0.18 $\pm$ 0.08 & --0.13 $\pm$ 0.06 & --0.23 $\pm$ 0.05 \\
RV UMa   & --1.20 $\pm$ 0.08 & 0.31 $\pm$ 0.07 & 0.25 $\pm$ 0.08 & 0.46 $\pm$ 0.11 & 0.22 $\pm$ 0.11 & --0.12 $\pm$ 0.11 & --0.20 $\pm$ 0.08 \\
UV Vir   & --1.10 $\pm$ 0.12 & 0.69 $\pm$ 0.10 & 0.41 $\pm$ 0.13 &  ...            & 0.10 $\pm$ 0.04 & --0.10 $\pm$ 0.08 &   0.04 $\pm$ 0.13 \\
\hline
\end{tabular}
\end{table*}

\subsection{Abundance ratios}
\label{sec:ratios}

\begin{figure}
\includegraphics[width=\columnwidth]{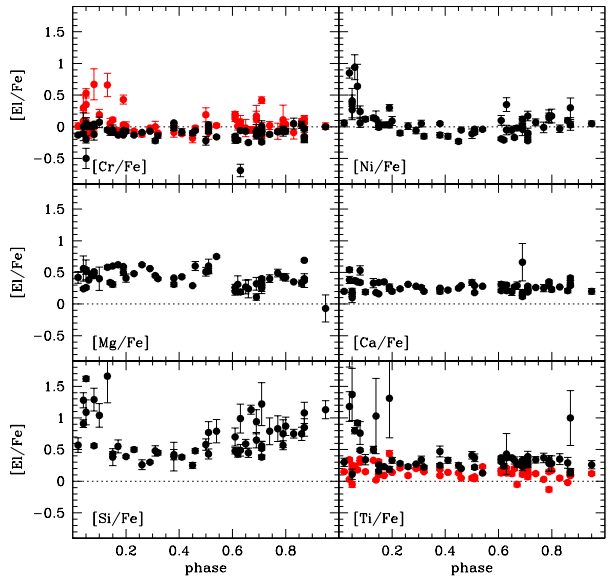}
\caption[]{Abundance ratios of single spectra plotted versus phase. Black dots
are for neutral species and red dots for ionized species (see text for
details), errorbars are the $\sigma$ of all surviving lines divided by the
square root of the number of lines.}
\label{fig:abophase}
\end{figure}

Abundance ratios for a few elements other than iron were computed for each
spectrum, providing uncertainties and final star weighted averages in the same
way used for iron (see Table~\ref{tab:ratios}). The behaviour of abundance
ratios with phase is illustrated in Figure~\ref{fig:abophase}. The resulting
abundance ratios are also plotted in Figure~\ref{fig:ratios} versus [Fe/H].

In summary, in spite of the obvious difficulties of obtaining parameters and
abundance ratios on variable stars spectra, especially when one observes with
long exposure times or outside the optimal phases, the results are stable and
broadly comparable to those obtained for non-variable stars {\em even rather
close to the shock phases}, if some extra effort is put into a careful selection
of spectral lines (we used an outlier rejection) and if one accepts the
inevitably higher errors and scatter (of the order of 0.1--0.15~dex, roughly
speaking) compared to non-variable stars. In other words, the
classical EW method based on static atmosphere models still works over most of
the pulsation cycle, including the early and main shock regions. However,
immediately after the main shock phase, we note a dangerous zone lying roughly
within $0\la\phi\la0.15$, depending on the species (see
Figure~\ref{fig:abophase}), which appears indeed to be the region where the
maximum disturbance on abundance determinations is reached. Our work fully
supports the findings by \citet{clementini95}, \citet{for11} or
\citet{wallerstein12}, and adds more insight into the main shock phase, where
indeed some neutral lines start disappearing \citep{chadid08}, but other lines
appear to remain reliable.

\subsubsection{Iron-peak elements} 

The iron-peak elements abundances of chromium and nickel
(Table~\ref{tab:ratios}), are based on approximately 5-15 well-behaved lines for
each element (see also Table~\ref{tab:ew} for a list of surviving lines),
depending on the spectrum. The scatter seems to increase for a few spectra
immediately after the main shock phase, and there are a few outliers, but when
all measurements for a star are combined through a weighted average, the
resulting abundance ratios are rather compatible with solar values, within the
highly varying uncertainties (Figure~\ref{fig:ratios}): the weighted averages
are $\langle$[Ni/Fe]$\rangle$=0.08$\pm$0.18~dex and
$\langle$[Cr/Fe]$\rangle$=--0.01$\pm$0.11~dex\footnote{[Cr/Fe] was obtained as
the weighted average of [Cr~I/Fe] and [Cr~II/Fe].}. Unlike \citet{for11}, we do
not find a large discrepancy between Cr~I and Cr~II: 
$\langle\Delta$[Cr/Fe]$\rangle$=--0.11$\pm$0.11~dex, but we observe that the
difference increases after the main shock phase (Figure~\ref{fig:abophase}) and
around and after the early shock phase, and that it goes in the same sense as
theirs \citep[see also][]{clementini95,sobeck07}. The two most metal-poor stars,
X~Ari and UZ~CVn, display rather large errorbars --- owing most certainly to the
paucity of lines --- but, for example, their enhanced nickel abundance is in
line with what found by \citet{gratton03} for field stars.

\subsubsection{$\alpha$-elements}

We also provide abundance ratios for Mg~I, Ca~I, Si~I, Ti~I, and Ti~II
(Table~\ref{tab:ratios}), based on approximately 3, 30, 10, 10, and 20 lines,
respectively (see Table~\ref{tab:ew}). Unlike the two iron-peak elements studied
above, calcium and magnesium do not seem to display any increase in the scatter
after the main shock, although there is a hint of a decrease in their abundance
before the shock \citep[see also][]{chadid08}. 

Silicon, which was extensively discussed by \citet{for11},
appears indeed extremely sensitive to the phase, with a larger spread at all
phases and very roughly following the T$_{\rm{eff}}$ trend. It was previously
reported in the literature \citep{shi09}, that silicon lines have largely
different NLTE effects and that bluer lines should have larger effects in
general with respect to redder lines. Our selection of lines (reported in
Table~\ref{tab:ew}), is based on the EW measurement quality and on statistical
rejections within the GALA routines, thus displaying a rough average effect,
with a large spread.

Apart from a few outliers, titanium behaves similarly to magnesium
and calcium; we can also observe a constant offset between Ti~I and Ti~II, of 
0.18$\pm$0.10~dex, with Ti~I higher than Ti~II. As discussed extensively by
\citet{for10} and \citet{for11}, who found a very similar result, the offset is
likely caused by uncertainities in the log$gf$ values of Ti~I lines, but lacking
a clearer explanation, we chose to obtain [Ti/Fe] as a weighted average of
[Ti~I/Fe] and [Ti~II/Fe], as done for Cr.

\begin{figure}
\includegraphics[width=\columnwidth]{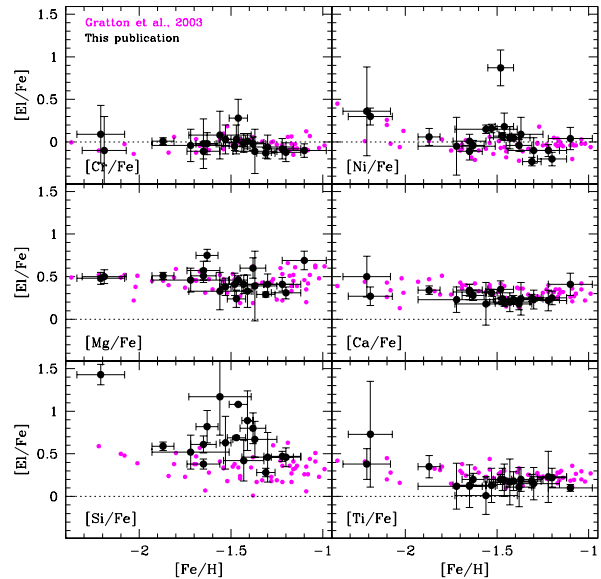}
\caption[]{Weighted average abundance ratios for each star with 1~$\sigma$ errors,
plotted versus [Fe/H] (see text for details). The field stars abundance ratios
measured by \citet{gratton03} are reported as magenta dots in all panels.}
\label{fig:ratios}
\end{figure}

Once the weighted averages of $\alpha$-element ratios were computed for each
star, we observed that the results were broadly compatible with the typical halo
$\alpha$-enhancement (Figure~\ref{fig:ratios}), except for silicon --- as said
above --- with no significant trends with [Fe/H]. We obtained:
$\langle$[Mg/Fe]$\rangle$=+0.43$\pm$0.12~dex,
$\langle$[Ca/Fe]$\rangle$=+0.27$\pm$0.10~dex,
$\langle$[Si/Fe]$\rangle$=+0.73$\pm$0.30~dex, and
$\langle$[Ti/Fe]$\rangle$=+0.26$\pm$0.23~dex. The average $\alpha$-enhancement
is $\langle$[$\alpha$/Fe]$\rangle$=+0.31$\pm$0.19~dex when excluding silicon
(when including it, both the enhancement and the error are higher by 0.1~dex).

\subsection{Literature comparisons}
\label{sec:lit2}

Five of our program stars were analyzed in the past with high-resolution
spectroscopy: TU~UMa, VX~Her, X~Ari, TW~Cap, and UY~Eri.

TU~Uma was observed at phase 0.77 by \citet{clementini95} and at phase 0.09 by
\citet{butler79}. Their adopted T$_{\rm{eff}}$ are fully consistent with the
shaded area in Figure~\ref{fig:teff}, covered by the \citet{for11} analysis, and
thus with our adopted value; their gravities are generally higher, more
consistent with the Baade-Wesselink results \citep[see Section~\ref{sec:grav}
and][]{for11}; their v$_{\rm{t}}$ are well above 3~km/s, i.e., 2--3.5~km/s
higher than our value, as expected because of the EW fitting method adopted here
(Section~\ref{sec:micro}). However, their resulting [Fe/H] abundances of
--1.05~dex \citep{butler79} and --1.55~dex \citep{clementini95} nicely bracket
our value of --1.31~dex, and our determination is compatible with both values,
considering their quoted errorbars of 0.1~dex \citep{clementini95} and 0.14~dex
\citep{butler79}.

VX~Her was observed by \citet{clementini95} at five phases in the range
0.54--0.69. Their T$_{\rm{eff}}$ value of 5950$\pm$115~K is included in the shaded
area in Figure~\ref{fig:teff}, and is thus compatible with our adopted value;
their log$g$ is 2.6~dex, again in line with the Baade-Wesselink determinations;
their v$_{\rm{t}}$ is 4.5~km/s, thus higher than our determination, as expected. Their
resulting [Fe/H]=--1.58~dex is very close to the one derived here of --1.56~dex.

The comparison with literature data for X~Ari is not straightforward. We found
three different high-resolution studies: \citet{clementini95};
\citet{lambert96}; and \citet{haschke12}. They obtained [Fe/H] of --2.47,
--2.47, and --2.61~dex, respectively, which are 0.3-0.4~dex lower than our
determination. These determinations are only marginally incompatible with the
one presented here, given our large combined uncertainty of 0.24~dex, that is
mostly due to our large uncertianties in the atmospheric parameters. As a test,
we ran again our abundance computations using fixed parameters, more in line
with the ones adopted in those studies, i.e., T$_{\rm{eff}}$=6300~K,
log$g$=2.5~dex and v$_{\rm{t}}$=4.5~km/s: we obtained a much lower [Fe/H]=--2.71~dex,
but at the expense of the ionization and excitation equilibria, and of the
flatness of the [Fe/H] vs. EW trend\footnote{More in detail, we obtain a slope
of --0.013~dex/eV for the [Fe/H] vs. excitation potential relation, a slope of
-0.16~dex/m\AA\  for the [Fe/H] vs. EW relation, and a difference of --0.18~dex
between Fe~I and Fe~II. With our parameters we obtain instead a slope of
0.0063~dex/eV for the [Fe/H] vs. excitation potential relation, a slope of
-0.44~dex/m\AA\  for the [Fe/H] vs. EW relation, and a difference of --0.11~dex
between Fe~I and Fe~II.}. We thus concluded that our spectrum of X~Ari does not
allow for such low metallicities and we proceeded to look for additional
literature determinations. We found: [Fe/H]=--2.01~dex with the $\Delta$S method
\citep{suntzeff94}; and [Fe/H]=--1.97~dex from the Baade-Wesselink method
\citep{blanco92}. These add to the [Fe/H]=--2.08~dex by \citet{kinemuchi06},
although \citet{beers00} reports --2.43~dex. Our result of --2.19~dex therefore
appears to lie in between two groups of discrepant literature values, and we
would weaken the internal consistency of our analysis if we adopted
significantly different atmospheric parameters. 

TW~Cap, one of our three Population II Cepheids, was studied by \citet{maas07}.
They do not report the phase of their observations, but they rejected all
spectra showing asymmetric line profiles, thus we can assume that their spectra
were far from the shock phases. Their parameters were obtained with a method
similar to ours, but they differ from our analysis, being:
T$_{\rm{eff}}$=5250~K, i.e., significantly lower than the typical estimates in
this paper and in \citet{for11}; log$g$=0.5~dex likewise among the lowest
estimates we encountered in the literature; v$_{\rm{t}}$=3.1~km/s, roughly
compatible with the \citet{for11} typical values, but higher than our typical
estimates, and lower than the typical v$_{\rm{t}}$ values in older abundance
analysis papers quoted so far. Finally, they obtain [Fe/H]=--1.8~dex, which is
formally consistent with our estimate.

UY~Eri is another of our Population II Cepheids, also studied by \citet{maas07}
with high-resolution spectroscopy. They obtain: T$_{\rm{eff}}$=6000~K, which is
substantially lower than our estimate; log$g$=1.5~dex, roughly compatible with
our estimate, v$_{\rm{t}}$=2.9~km/s, higher than our estimate. In this case,
their final [Fe/H]=--1.84~dex does not agree with our derived
--1.43$\pm$0.20~dex. We thus tried employing the parameters adopted by
\citet{maas07}, even if we do not know at which pulsation phase they were
evaluated, and examined the impact on our abundance analysis. As expected, the
constancy of [Fe/H] versus excitation potential, EW, and wavelength were totally
disrupted, together with the ionization balance between Fe~I and Fe~II, thus
proving that our data cannot support parameters much different than the ones
adopted here. However, the resulting [Fe/H] went substantially down, below
--2~dex, showing that the different choice of the parameters is the most likely
cause for the different iron abundances. 

In conclusion, the presented literature comparisons (see also
Section~\ref{sec:iron}) generally support the results obtained with our data and
method. There are marginal disagreements for X~Ari and UY~Eri, that can be at
least partially explained by the different assumptions on atmospheric
parameters, as discussed above. 

\section{Discussion and conclusions}
\label{sec:disandcon}

We have analysed a sample of 21 variable stars --- mostly RRab, with one RRc,
two BL~Her and one W~Vir --- having high-resolution spectra from both
proprietary (UVES and SARG) and archival (UVES, HARPS, FEROS, APO) sets. The
data were taken at random phases and as a consequence several spectra were
obtained outside optimal phases, and a few spectra very close to the shock
phases. 

We performed a classical EW-based abundance analysis with static stellar
atmospheric models and obtained consistent metallicities
(Section~\ref{sec:iron}) and abundance ratios (Sections~\ref{sec:ratios}) ---
within roughly 0.15~dex --- regardless of the phase, even for those spectra very
close to the shock phases. It is interesting to note that the
W~Vir variable (TW~Cap) and the two BL~Her variables (UY~Eri and V716~Oph)
behave exactly like RRab at all phases, and thus appear virtually
indistinguishable, from the spectroscopic analysis point of view, at least in
the sampled phases. While this result is not surprising, given that the physical
mechanism responsible for the pulsation is the same and the stellar parameters
are quite similar, this is the first time a sample comprising both RR~Lyrae
(RRab and RRc) and Population II Cepheids is analyzed homogeneously.

Comparisons with the \citet{beers00} and \citet{kinemuchi06} samples, based on
[Fe/H] calibrations for RR~Lyrae, showed a rather satisfactory scatter of
roughly 0.20~dex, with no significant offset or phase trend
(Section~\ref{sec:iron}), and comparisons with high-resolution studies
(Section~\ref{sec:lit2}) generally supported our abundance analysis. Iron-peak
elements were overall solar, again with no trends with phase and a moderate
scatter of roughly 0.15~dex. An average [$\alpha$/Fe]=+0.31$\pm$0.19~dex was
found over the entire sample, based on Mg, Ca, Ti~I, and Ti~II. The expected
problematic element was silicon, which was discussed in detail in the literature
\citep[see][for  references]{for11}: it was rather difficult to find reliable
lines, producing relatively small scatter and a [Si/Fe] abundance independent
from the phase. Also, a small but systematic ionization imbalance of
$\langle\Delta$[Cr/Fe]$\rangle$=--0.11$\pm$0.11~dex and
$\langle\Delta$[Ti/Fe]$\rangle$=+0.18$\pm$0.10~dex was observed, as reported by
previous authors, and most probably caused by uncertainties in the log$gf$
values. 

Our spectroscopically derived atmospheric parameters are also broadly consistent
with the ones expected from template curves (Sections~\ref{sec:teff},
\ref{sec:grav}, and \ref{sec:micro}) for RRab and Population~II Cepheids, while
for U~Com, the only RRc in our sample, template curves give a too high
T$_{\rm{eff}}$ that was not reported in previous work \citep[see for
example][]{lambert96,govea14}. Also, the atmospheric parameters found in the
present analysis were broadly consistent with the ones found in the similar
analysis performed by \cite{for11}, with scatters around the expected values of
roughly $\pm$300~K for T$_{\rm{eff}}$, $\pm$0.3~dex for $\log g$, and
$\pm$0.4~km/s for v$_{\rm{t}}$.

In summary, a classical EW-based abundance analysis on high-resolution
(R$\geq$30\,000), high S/N spectra (S/N$\geq$30) is appropriate to study
RR~Lyrae spectra at all phases, with a possible danger only in
the range $0\la\phi\la0.15$ (where the abundance ratios could be overestimated,
depending on the species), and with relatively long exposure times (up to
45~min), provided absorption lines are carefully selected and as long as an
error of about 0.10--0.15~dex is considered acceptable.

\section*{Acknowledgments}

We would like to thank E.~Bernard for useful discussions on the [Fe/H]
calibrations of RR~Lyrae based on pulsation properties, and G.~Fiorentino for
help on the Population~II Cepheids properties and classification.
We warmly thank C.~Sneden, who was the referee of this paper,
for his insightful comments. N.~Britavskiy acknowledges the support of the
European Commission Erasmus Mundus LOT~7 programme and warmly thanks the
INAF\,--\,Bologna Observatory, where most of this work was carried out, for the
hospitality during the grant stay. D.~Romano acknowledges financial support from
PRIN MIUR 2010--2011, project ``The chemical and dynamical evolution of the
Milky Way and Local Group galaxies", prot. 2010LYSN2T. This publication makes
use of data products from the Two Micron All Sky Survey, which is a joint
project of the University of Massachusetts and the Infrared Processing and
Analysis Center/California Institute of Technology, funded by the National
Aeronautics and Space Administration and the National Science Foundation. This
publication makes use of the data from the Northern Sky Variability Survey
(NSVS) created jointly by the Los Alamos National Laboratory and University of
Michigan. The NSVS was funded by the Department of Energy, the National
Aeronautics and Space Administration, and the National Science Foundation. This
research made use of the products of the Cosmic-Lab project funded by the
European Research Council. In this work we made extensive use of the NASA ADS
abstract service, of the Strasbourg CDS database, and of the atomic data
compiled in the VALD data base.

\label{lastpage}
\end{document}